\documentclass[prd,aps,preprint,showkeys,showpacs,preprintnumbers]{revtex4-1}
\usepackage{physics,slashed}
\usepackage[retainorgcmds]{IEEEtrantools}
\usepackage[b]{esvect}
\usepackage{footnote}
\usepackage{hyperref}
\usepackage{multirow}
\usepackage{graphicx}
\newcommand{\sld}{\slashed}

\newcommand{\dlt}[2]{\delta^{(#1)}(#2)}
\newcommand{\frao}[1]{\frac{1}{#1}}

\newcommand{\gma}{\gamma}

\newcommand{\napm}{\!\pm\!}
\newcommand{\IEyn}{\IEEEyesnumber}
\newcommand{\IEnn}{\IEEEnonumber}
\newcommand{\IEysn}{\IEEEyessubnumber}
\newlength{\IEEEglue}
\IEEEeqnarraydefcolsep{0}{\IEEEglue}
\begin{document}
\title{Decay Constants and Distribution Amplitudes of B Meson in the
  Relativistic Potential Model}
\author{Hao-Kai Sun}
\email{sunhk@mail.nankai.edu.cn}
\author{Mao-Zhi Yang}
\email{yangmz@nankai.edu.cn}
\affiliation{School of Physics,
  Nankai University, Tianjin 300071, People's Republic of China}
\date{\today}
\begin{abstract}
  In this work we study the decay constants of $B$ and $B_s$ mesons
  based on the wave function obtained in the relativistic potential
  model. Our results are in good agreement with experimental data which
  enables us to apply this method to the investigation of $B$-meson
  distribution amplitudes. A very compact form of the distribution
  amplitude is obtained. We also investigate the one-loop QCD
  corrections to the pure leptonic decays of $B$ mesons. We find that,
  after subtracting the infrared divergence in the one-loop
  corrections using the factorization method, the QCD one-loop
  corrections to the hard amplitude of  leptonic decay  will be zero.
\end{abstract}
\pacs{12.39.Pn,12.39.St,13.30.Ce,14.40.Nd}
\keywords{Decay Constant, Distribution Amplitudes, Factorization}
\maketitle
\section{Introduction}
\label{sec:intro}
The study of $B$-meson decays, especially the exclusive semileptonic
and two-body nonleptonic decays, presents rich information for testing
and understanding the standard model (SM). In the past two decades, as
the running and upgrading of $B$-factories, a great amount of
experimental data has been accumulated. Although a lot of models
and/or approaches have been developed in theory, the poor knowledge of
nonperturbative quantum chromodynamics (QCD) effects still limits
theoretical predictions severely. In two-body nonleptonic decays of
$B$-meson, QCD factorization
\cite{Beneke:1999br,Beneke:2000ry,Beneke:2001ev,Beneke:2003zv} and
perturbative QCD approaches
\cite{Keum:2000ph,Keum:2000wi,Lu:2000em,Keum:2002cr,Keum:2003js} have
been developed, which allow us to separate the nonperturbative effect
out as universal quantities, such as, the light-cone distribution
amplitudes (LCDA) and/or form factors. The $B$-meson LCDA has been
studied extensively. Several forms of the distribution amplitudes are
proposed or obtained by some theoretical methods such as solving the
equations of motion in the literature
\cite{Grozin:1996pq,Beneke:2000wa,Lange:2003ff,Lee:2005gza,Kawamura:2001jm,Huang:2004na,Huang:2005kk,Hwang:2010hw,Bell:2013tfa}.

Inspired by the construction of initial bound state in
Ref.\cite{Leutwyler:1984je} and based on our previous works on the
mass spectrum and wave functions of $B$-meson
\cite{Yang:2011ie,Liu:2013maa,Liu:2015lka}, we try an alternate way
to study the distribution amplitudes with the help of wave functions
obtained in the relativistic potential model
\cite{Liu:2013maa,Liu:2015lka}. Considering the recent experimental
data on the pure leptonic decays of $B$ mesons,  we
focus on a careful investigation about the decay constants and the
distribution amplitudes (DAs) of $B$-mesons in this paper.

In general, the decay constants of charged heavy-light mesons are
related directly to the pure leptonic decay widths and thus measuring
decay constants can provide a chance to check different theoretical
models and may also give some hints for physics beyond the standard
model (SM). During the past decades, many methods have been applied to
the study of the decay constants, such as, QCD sum rules
\cite{Penin:2001ux,Bordes:2004vu,Bordes:2005wi,Lucha:2011zp,Narison:2012xy,Gelhausen:2013wia,Narison:2015nxh},
the Bethe-Salpeter equation \cite{Wang:2004xs,Cvetic:2004qg}, the
field correlator method \cite{Badalian:2007km}, the soft-wall
holographic approach \cite{Branz:2010ub}, the potential models
\cite{Godfrey:1985xj,Colangelo:1990rv,DiPierro:2001dwf,Ebert:2006hj,Yang:2011ie},
and the lattice QCD simulations
\cite{Davies:2003ik,Gray:2005ad,Davies:2010ip,McNeile:2011ng,Bazavov:2011aa,Becirevic:2012ti,Na:2012kp,Dowdall:2013tga},
etc. Up to now there are still large uncertainties for the value of
$\abs{V_{ub}}$ \cite{Agashe:2014kda}, and only the pure leptonic decay
mode of $B$ meson with $\tau$ lepton in the final state has been
measured in experiment
\cite{Adachi:2012mm,Kronenbitter:2015kls,Lees:2012ju,Aubert:2009wt}
(also with large uncertainties). Our result for the branching ratio of
$B\to\tau \nu $ decay is well located in the experimental error bars
\cite{Adachi:2012mm,Kronenbitter:2015kls,Lees:2012ju,Aubert:2009wt,Rosner:2015wva}.
Further tests from experiments are needed in the future with enhanced
precision (most possibly come from the Belle II / SuperKEKB
collaboration \cite{Barrett:2015gvz,Wang:2015kmm}).

We study the $B$-meson distribution amplitudes in this work. The
analytical forms both in coordinate and momentum space are obtained.
When they are transformed to the commonly used form of LCDA, the
figures show that they obey the model-independent limitations
\cite{Lee:2005gza}. We also consider the pure leptonic decays of
$B$-meson up to one-loop level in QCD corrections. We find that
one-loop corrections to the hard-scattering kernel in QCD will be zero
after subtracting the infrared divergence by using the factorization
method.

The paper is organized as followings. In Sec.\ref{sec:dc}, we
calculate the decay constants of the $B$ and $B_s$ mesons. The
branching ratios of leptonic decays of $B$ meson are also calculated
and compared with experimental data. In Sec.\ref{sec:da}, the matrix
element between $B$ meson and vacuum state, which defines the
distribution amplitudes (DAs), is studied. The analytical form of the
matrix element and DAs are obtained and figures are shown as
illustrations. We finally obtain a compact expression for the matrix
element. Section \ref{sec:qcdf} is devoted to the study of the pure
leptonic decay of the $B$-mesons up to one-loop level in QCD and
Sec.\ref{sec:concl-or-disc} is for the conclusion and discussion.
\section{Decay Constants of $B$ and $B_s$ Mesons}
\label{sec:dc}
Recently, the spectra of heavy-light quark-antiquark system have been
studied in the relativistic potential model in our previous works
\cite{Yang:2011ie,Liu:2013maa,Liu:2015lka}, where hyperfine
interactions are included \cite{Liu:2013maa,Liu:2015lka}. The whole
spectra of $B$ and $D$ system are in well agreement with experimental
measurements. Hence in this work, we extend our previous works
\cite{Liu:2013maa,Liu:2015lka} by studying the decay properties of $B$
meson with the wave functions obtained in the relativistic potential
model. We study the decay constants of $B$ and $B_s$ mesons at first,
and then give a compact form of distribution amplitudes of $B$-meson,
which shall be useful for studying $B$ decays.

The decay constant of a pseudoscalar meson is defined by the matrix
element of the axial current between the meson and the vacuum state
\begin{IEEEeqnarray}{cx}
  \mel{0}{\bar{q}\gma^{\mu}\gma^5 Q}{P}=if_PP^{\mu}\label{eq:1}
\end{IEEEeqnarray}
where the axial current is composed of a light antiquark field
$\bar{q}$ and a heavy quark field $Q$.

The pseudoscalar meson as a bound state of a quark and antiquark
system can be described by ~\cite{Leutwyler:1984je,Yang:2011ie},
\begin{IEEEeqnarray}{rl}
  \ket{P(\vv{P})}=\frao{\sqrt{N_L}}\frao{\sqrt{3}}
  \sum_i&\int\!\dd[3]{k_q}\dd[3]{k_Q}\dlt{3}{\vv{P}-\vv{k}_q
  -\vv{k}_Q}\varPsi_0(\vv{k}_q)\IEnn\\
  &\frao{\sqrt{2}}\qty[c^{i\dagger}(\vv{k}_Q,\uparrow)
  b^{i\dagger}(\vv{k}_q,\downarrow)-
  c^{i\dagger}(\vv{k}_Q,\downarrow)
  b^{i\dagger}(\vv{k}_q,\uparrow)]
  \ket{\stackrel{\phantom{\rightarrow}}{0}}\label{eq:2}
\end{IEEEeqnarray}
where $N_L$ is the normalization factor, and the normalization
conditions will be shown explicitly below. $i$ stands for the QCD
color index and $\frao{\sqrt{3}}$ is the corresponding normalization
factor. The factor $\frao{\sqrt{2}}$ is the normalization factor for
the quark spin states which are indexed by up or down arrows. Inside
the square parenthesis, $b^{i\dagger}$ and $c^{i\dagger}$ are the
creation operators of the light antiquark $\bar{q}$ and the heavy
quark $Q$, respectively.

The function $\varPsi_0(\vv{k}_q)$ is the normalized wave function of
the pseudoscalar meson at ground state in the momentum space, which
describes the wave function of the quark and antiquark constituents in
a meson. It is noted here that these quark constituents are the
effective quarks carrying a gluon cloud and therefore the quarks have
constituent masses~\cite{Wirbel:1985ji}.

The wave function can be obtained by solving the Schr\"odinger type wave
equation with relativistic dynamics
\begin{IEEEeqnarray}{cx}
  (H_0+H')\varPsi(\vv{r})=E\varPsi(\vv{r}),\label{eq:3}
\end{IEEEeqnarray}
where $H_0+H'$ is the effective Hamiltonian (its explicit
expression can be found in Ref.~\cite{Liu:2015lka}) and $E$ is the
energy of the meson. The first term $H_0$ contains the kinetic part
and the effective potential which is taken as a combination of a
Coulomb term and a linear confining term inspired by
QCD~\cite{Godfrey:1985xj,Eichten:1978tg,Eichten:1979ms}.

The second term $H'$ is the spin-dependent part of the Hamiltonian
including contributions of one-gluon-exchange diagram in the
nonrelativistic approximation \cite{DeRujula:1975smg,Godfrey:1985xj}
and new terms which account for contributions of nonperturbative
dynamics in the bound state system and relativistic corrections for
the light quark in the heavy meson~\cite{Liu:2013maa,Liu:2015lka}.

The normalization conditions for wave function are
\begin{IEEEeqnarray}{cx}
  \int\dd[3]{k}\abs{\varPsi_0(\vv{k})}^2=1,
  \IEyn\IEysn\label{eq:4a}\\
  \qty{c(\vv{k},s),c^{\dagger}(\vv{k}',s')}
  =\delta_{ss'}\dlt{3}{\vv{k}-\vv{k}'},\IEysn\label{eq:4b}\\
  \ip{P(\vv{P})}{P(\vv{P}')}=(2\pi)^32E
  \dlt{3}{\vv{P}-\vv{P}'}.\IEysn\label{eq:4c}
\end{IEEEeqnarray}
Note that we omit the color index of the operator $c$ and use $s,s'$
to denote the spin states. Substituting Eq.\eqref{eq:2} into
Eq.\eqref{eq:4c} and using Eq.\eqref{eq:4a} and Eq.\eqref{eq:4b}, we
can obtain the normalization factor
\begin{IEEEeqnarray}{cx}
  N_L=\frao{(2\pi)^32E}.\label{eq:5}
\end{IEEEeqnarray}

\begin{figure}[tbp]
  \centering
  \includegraphics[width=0.5\textwidth]{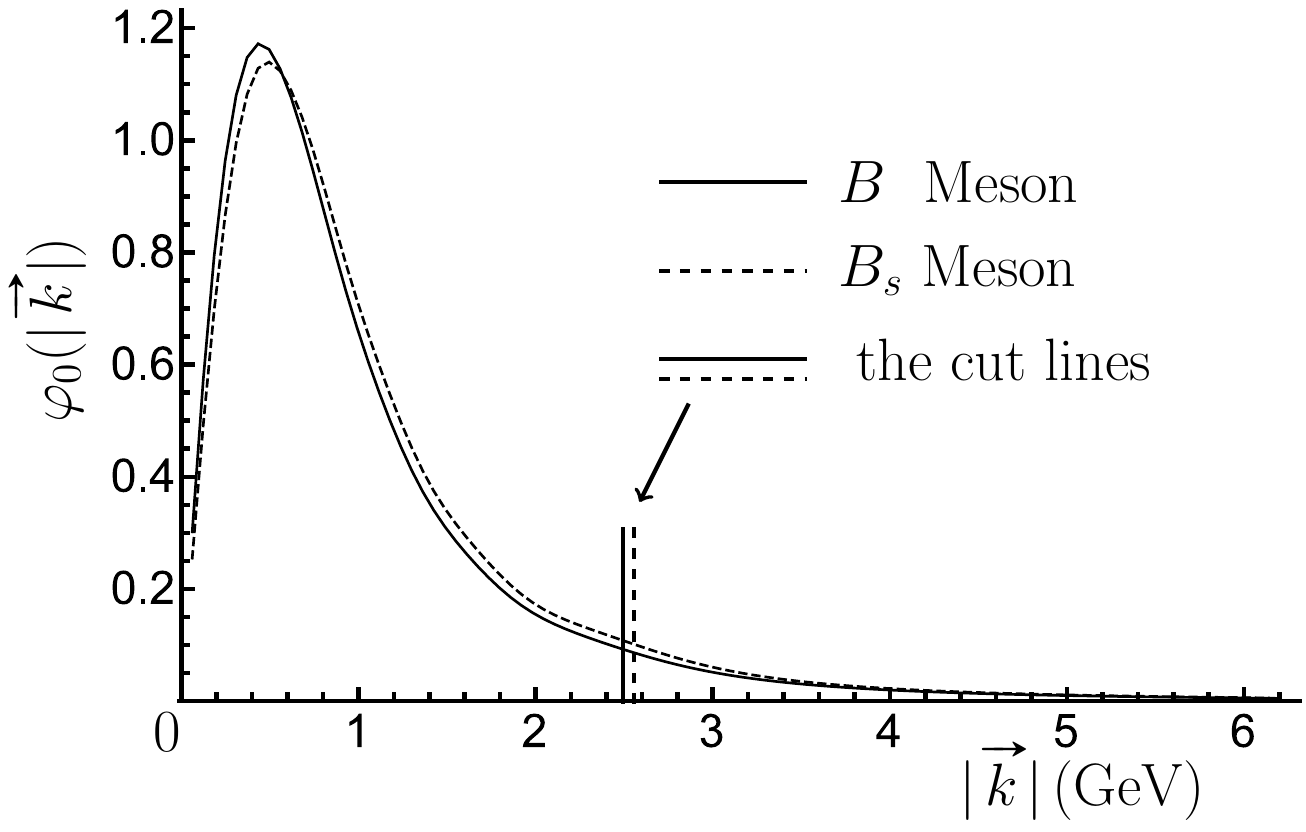}
  \caption{\label{fig:1} Reduced wave functions for $B$-meson.}
\end{figure}

The wave function has been solved numerically in our previous work
\cite{Liu:2015lka}. For $B$ meson, the wave function can be expressed
by
\begin{IEEEeqnarray}{cx}
  \varPsi_0(\vv{k})=
  \frac{\varphi_0(|\vv{k}|)}{|\vv{k}|} Y_{00}(\theta,\phi)
  \label{eq:6}
\end{IEEEeqnarray}
where $\varphi_0(|\vv{k}|)$ is the reduced wave function. The
numerical result of $\varphi_0(|\vv{k}|)$ can be shown in
Fig.\ref{fig:1}.

\begin{figure}[tbp]
  \centering
  \includegraphics[width=\textwidth]{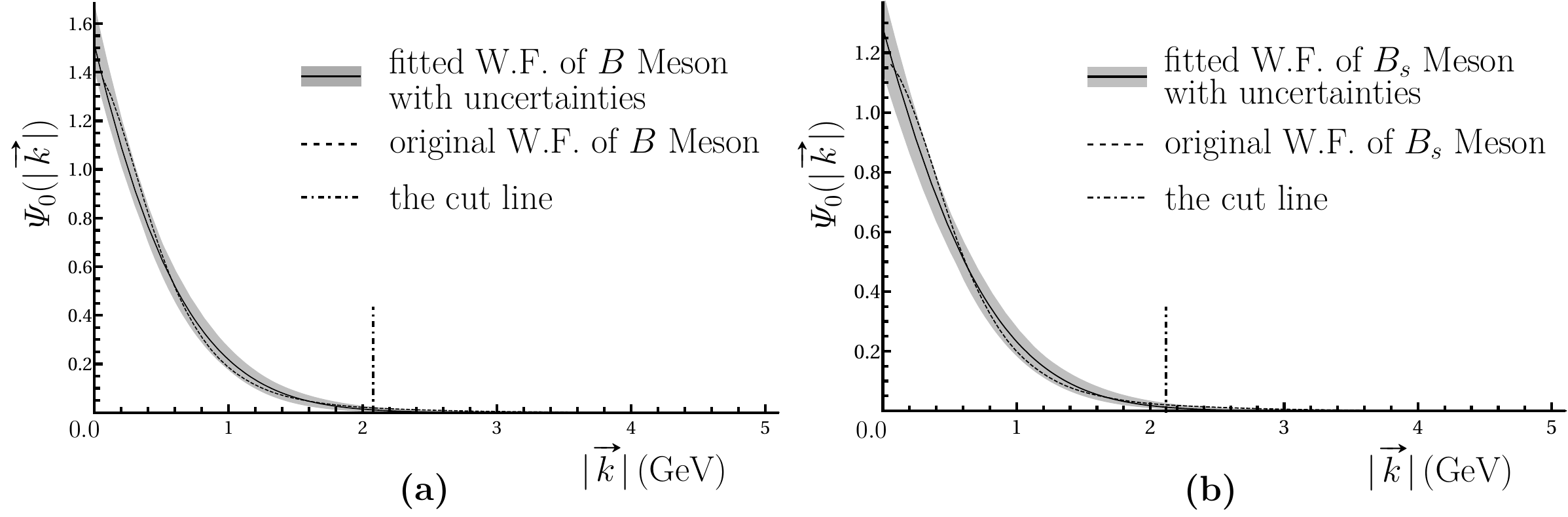}
  \caption{\label{fig:2}The wave functions (W.F.) of $B$-meson.}
\end{figure}

Since it is convenient to have an analytical form of the wave function
$\varPsi_0(\vv{k})$ for the numerical calculation, we fit the wave
function obtained in our previous work \cite{Liu:2015lka} with an
exponential function and finally obtain the fitted form for the
$B_{(s)}$ meson wave function with combined theoretical uncertainties
as
\begin{IEEEeqnarray}{cx}
  \varPsi_0(\vv{k})=
  a_1e^{a_2|\vv{k}|^2+a_3|\vv{k}|+a_4}\label{eq:7},
\end{IEEEeqnarray}
where the parameters including uncertainties for $B$ meson are
\begin{IEEEeqnarray}{llx}
  a_1=4.55_{-0.30}^{+0.40}\,\mathrm{GeV}^{-3/2},\quad&
  a_2=-0.39_{-0.20}^{+0.15}\,\mathrm{GeV}^{-2};\nonumber\\
  a_3=-1.55\!\pm\!0.20\,\mathrm{GeV}^{-1},\quad&
  a_4=-1.10_{-0.05}^{+0.10},
\end{IEEEeqnarray}
and for $B_s$ meson:
\begin{IEEEeqnarray}{llx}
  a_1=1.60_{-0.18}^{+0.15}\,\mathrm{GeV}^{-3/2},\quad&
  a_2=-0.43_{-0.10}^{+0.15}\,\mathrm{GeV}^{-2};\nonumber\\
  a_3=-1.28_{-0.20}^{+0.18}\,\mathrm{GeV}^{-1},\quad&
  a_4=-0.22_{-0.08}^{+0.06}.
\end{IEEEeqnarray}
The uncertainties for the parameters ensure that the deviation of the
wave function from its central value is approximately about 8\%. The
illustrations for the fit of the wave functions are shown in
Fig.\ref{fig:2}, where the grey bands denote the relevant
uncertainties for the wave functions of $B$ and $B_s$ mesons.

In the calculation of the decay constants, four-momentum conservation
should hold
\begin{IEEEeqnarray}{cx}
  k_q+k_Q=P,\label{eq:8}
\end{IEEEeqnarray}
where $k_{q,Q}$ and $P$ are the momenta of the quark constituents and
the meson, respectively.

With the restriction above, we consider the ACCMM scenario
\cite{Altarelli:1982kh,Colangelo:1998eb}, where the light quark is
kept on-shell, while the heavy quark off-shell,
\begin{IEEEeqnarray}{cx}
  E_q+E_Q=m_P,\IEyn\IEysn\label{eq:9a}\\
  E_q^2=m_q^2+|\vv{k}|^2,\IEysn\label{eq:9b}\\
  m_Q^2(\vv{k})=E_Q^2-|\vv{k}|^2.\IEysn\label{eq:9c}
\end{IEEEeqnarray}
Equation \eqref{eq:9a} is the energy conservation in the meson rest frame.
We assume that the running mass of the heavy quark must be positive
$m_Q(\vv{k})\geq 0$. Thus the actual range of the momentum $|\vv{k}|$
is restricted under a particular value, which is shown as the cut
lines in Figs.\ref{fig:1} and \ref{fig:2}.

Substituting Eq.\eqref{eq:2} into Eq.\eqref{eq:1} in the rest frame
and contracting the quark (antiquark) creation operators with the
annihilation operators in the quark field of the axial current
$\bar{q}\gma^{\mu}\gma^5Q$, we obtain
\begin{IEEEeqnarray}{cx}
  f_P=\sqrt{\frac{3}{(2\pi)^3m_P}}\int\!\dd[3]{k}\varPsi_0(\vv{k})
  \frac{\qty(E_q+m_q)\qty(E_Q+m_Q)-|\vv{k}|^2}
  {\sqrt{E_qE_Q\qty(E_q+m_q)\qty(E_Q+m_Q)}},\label{eq:10}
\end{IEEEeqnarray}
where the integral over the variable $\vv{k}$ should be limited in
the finite range according to Eqs.\eqref{eq:9a}--\eqref{eq:9c}.

The parameters used in this work are \cite{Liu:2015lka}
\begin{IEEEeqnarray}{ll}
  m_s=0.32\,\mathrm{GeV}\qc&m_u=m_d=0.06\,\mathrm{GeV},\;\;
  m_b=4.99\,\mathrm{GeV},\label{eq:11}
\end{IEEEeqnarray}
and the mesons' masses are taken from PDG~\cite{Agashe:2014kda}
\begin{IEEEeqnarray}{cx}
  m_B=5.28\,\mathrm{GeV}\qc m_{B_s}=5.37\,\mathrm{GeV}.\label{eq:12}
\end{IEEEeqnarray}

The errors are estimated by varying the parameters in the allowed
ranges. The total errors are around 7\% for the decay constants of $B$
and $B_s$ mesons. We also calculate the ratio of the decay constants
of $B$ and $B_s$ mesons $f_{B_s}/f_B$. The final results obtained are
\begin{IEEEeqnarray}{cx}
  f_B=219\pm 15\,\mathrm{MeV}\qc
  f_{B_s}=266\pm 19\,\mathrm{MeV}\qc
  f_{B_s}/f_B=1.21\pm 0.09\,.\label{eq:13}
\end{IEEEeqnarray}

During past decades, many theoretical methods or models have been
developed for the calculation of the $B$-meson decay constants. In
this paper, we list some of the results for comparison in
Table.\ref{tab:i}, where one can see that our results are consistent
with most of the theoretical predictions.

\begin{table}[tbp]
\centering
\begin{tabular}{l@{\quad}l@{\quad}l@{\quad}l@{\quad}l}
  \hline
  \hline
  \textbf{Reference}&\textbf{Method}
  &$f_B$\,(MeV)&$f_{B_s}$\,(Mev)&$f_{B_s}/f_B$\\
  \hline
  this work&RPM$^{\ast}$
  &$219\pm 15$&$266\pm 19$&$1.21\pm 0.09$\\
  Colangelo 91 \cite{Colangelo:1990rv}&RPM
  &$230\pm 35$&$245\pm 37$&$1.07\pm 0.17$\\
  Cveti\u{c} 04 \cite{Cvetic:2004qg}&QM BS$^{\ddag}$
  &$196\pm 29$&$216\pm 32$&$1.10\pm 0.18$\\
  Badalian 07 \cite{Badalian:2007km}&FCM$^{\pounds}$
  &$182\pm 8$&$216\pm 8$&$1.19\pm 0.03$\\
  Hwang 09 \cite{Hwang:2009qz}&LFQM$^{\S}$
  &$204\pm 31$&$270.0\pm 42.8$&$1.32\pm 0.08$\\
  HPQCD 11 \cite{McNeile:2011ng}&LQCD (2+1)$^{\P}$
  &\qquad --&$225\napm 3\napm 3$&\qquad --\\
  FNAL/MILC 11 \cite{Bazavov:2011aa}&LQCD (2+1)
  &$196.9\napm 5.5\napm 7.0$&$242.0\napm 5.1\napm 8.0$
  &$1.229\napm 0.013\napm 0.023$\\
  HPQCD 12 \cite{Na:2012kp}&LQCD (2+1)
  &$191\napm 1\napm 8$&$228\napm 3\napm 10$
  &$1.188\napm 0.012\napm 0.013$\\
  Narison 12 \cite{Narison:2012xy}&QCD SR$^{\dag}$
  &$206\pm 7$&$234\pm 5$&$1.14\pm 0.03$\\
  Gelhausen 13 \cite{Gelhausen:2013wia}&QCD SR
  &$207_{-9}^{+17}$&$242_{-12}^{+17}$&$1.17_{-0.04}^{+0.03}$\\
  HPQCD 13 \cite{Dowdall:2013tga}&LQCD (2+1+1)
  &$184\pm 4$&$224\pm 5$&$1.217\pm 0.008$\\
  ETM 13 \cite{Carrasco:2013naa}&LQCD (2+1+1)
  &$196\pm 9$&$235\pm 9$&$1.201\pm 25$\\
  Aoki 14 \cite{Aoki:2014nga}&LQCD (2+1)
  &$218.8\napm 6.4\napm 30.8$&$263.5\napm 4.8\napm 36.7$
  &$1.193\napm 0.020\napm 0.044$\\
  RBC/UKQCD 14 \cite{Christ:2014uea}&LQCD (2+1)
  &$195.6\napm 6.4\napm 13.3$&$235.4\napm 5.2\napm 11.1$
  &$1.223\napm 0.014\napm 0.070$\\
  Wang 15 \cite{Wang:2015mxa}&QCD SR
  &$194\pm 15$&$231\pm 16$&$1.19\pm 0.10$\\
  \hline
  \hline
\end{tabular}
\\\vspace{3mm}
\begin{minipage}{1.0\linewidth}
  \begin{itemize}
    {\footnotesize
    \item[$^{\ast}$] Relativistic potential model. \\\vspace{-5mm}
    \item[$^{\dag}$] QCD sum rules. \\\vspace{-5mm}
    \item[$^{\ddag}$] Quark model based on Bethe-Salpeter equation. \\\vspace{-5mm}
    \item[$^{\P}$] lattice-QCD with dynamical quark flavors $N_f$ in
      the parentheses. \\\vspace{-5mm}
    \item[$^{\S}$] Light-front quark model. \\\vspace{-5mm}
    \item[$^{\pounds}$] Field correlator method.
    }
  \end{itemize}
\end{minipage}
\caption{\label{tab:i}Theoretical results of the decay constants of $B$-mesons.}
\end{table}

The branching ratio of the leptonic decay of $B$ meson can be
calculated by the following formula
\begin{IEEEeqnarray}{cx}
  \mathcal{B}\qty(B^{\pm}\rightarrow l^{\pm}\nu)=\frac{G_F^2m_l^2m_B}{8\pi}
  \qty(1-\frac{m_l^2}{m_B^2})^2f_B^2\abs{V_{ub}}^2\tau_B,\label{eq:14}
\end{IEEEeqnarray}
where $G_F$ is the Fermi constant, $V_{ub}$ the
Cabibbo-Kobayashi-Maskawa (CKM) matrix element, $m_B$ and $m_l$ the
masses of $B^{\pm}$ meson and lepton, respectively, and $\tau_B$ is
the life time of $B^{\pm}$ meson.

In this work, we obtain
\begin{IEEEeqnarray}{rCl}
 \mathcal{B}\qty(B^+\rightarrow e^+\nu_e)&=
 &\qty(1.17\pm 0.18)\times 10^{-11},\IEyn\IEysn\label{eq:15a}\\
 \mathcal{B}\qty(B^+\rightarrow \mu^+\nu_{\mu})&=
 &\qty(5.01\pm 0.78)\times 10^{-7},\IEysn\label{eq:15b}\\
 \mathcal{B}\qty(B^+\rightarrow \tau^+\nu_{\tau})&=
 &\qty(1.41\pm 0.22)\times 10^{-4},\IEysn\label{eq:15c}
\end{IEEEeqnarray}
where the errors mainly come from the uncertainties of the decay
constants $f_B$ and the CKM matrix element
$|V_{ub}|$~\cite{Agashe:2014kda}
\begin{IEEEeqnarray}{cx}
  |V_{ub}|=\qty(4.09\pm 0.39)\times 10^{-3}.\label{eq:16}
\end{IEEEeqnarray}

The branching ratio of $B\rightarrow\tau^+\nu_{\tau}$ channel has been
measured by Belle and \emph{BABAR} collaborations
\cite{Adachi:2012mm,Kronenbitter:2015kls,Lees:2012ju,Aubert:2009wt}.
The results are shown in Table \ref{tab:ii}.
\begin{table}[tbp]
\centering
\begin{tabular}{|c|c|l|}
  \hline
  \textbf{Experiment}&\textbf{Tag}
  &$\mathcal{B}$(units of $10^{-4}$)\\
  \hline
  Belle\cite{Adachi:2012mm}&Hadronic
  &\rule{0pt}{3ex}$0.72_{-0.25}^{+0.27}\pm 0.11$\\
  Belle\cite{Kronenbitter:2015kls}&Semileptonic
  &\rule{0pt}{3ex}$1.25\pm 0.28\pm 0.27$\\
  \hline
  \emph{BABAR}\cite{Lees:2012ju}&Hadronic
  &\rule{0pt}{3ex}$1.83_{-0.49}^{+0.53}\pm 0.24$\\
  \emph{BABAR}\cite{Aubert:2009wt}&Semileptonic
  &\rule{0pt}{3ex}$1.7\pm 0.8\pm 0.2$\\
  \hline
\end{tabular}
\caption{\label{tab:ii}Experimental results for $\mathcal{B}(B^+\rightarrow
  \tau^+\nu_{\tau})$.}
\end{table}

Taking the large uncertainties of the experimental data into
consideration, our predicted branching ratio of the decay channel
$B^+\rightarrow\tau^+\nu_{\tau}$ [Eq.\eqref{eq:15c}] is consistent
with the experimental results.

As an upgrade of the Belle / KEKB experiment, the Belle II / SuperKEKB
will start taking data from 2018. With a designed luminosity $8\times
10^{35}\,\mathrm{cm}^{-2}\mathrm{s}^{-1}$, which is about 40 times
larger than its predecessor, data sample corresponding to
$50\,\mathrm{ab}^{-1}$ will be accumulated within five years of
operation~\cite{Wang:2015kmm}. It is expected to reduce both the
statistical and systematic errors of the
$B^+\rightarrow\tau^+\nu_{\tau}$ decay mode by a factor about
7~\cite{Pakhlov:2016yzk}.
\section{$B$-Mesons Distribution Amplitudes}
\label{sec:da}
Based on the success of our predictions on the mass spectra
\cite{Yang:2011ie,Liu:2013maa,Liu:2015lka} and the decay constants of
$B$-mesons, we continue to study of the matrix element of $B$ meson
which defines the DAs. The matrix element and DAs are generally used
in studying hadronic decays of $B$ meson.

Generalizing the current in the definition of the decay constant in
Eq.\eqref{eq:1} from local to nonlocal operators and making use of
Fierz identity, we obtain the matrix element between the $B$ meson and
the vacuum state in coordinate space
\begin{IEEEeqnarray}{rCl}
  \tilde{\Phi}_{\alpha\beta}(z)&\equiv&\mel{0}
  {\bar{q}_{\beta}(z)[z,0]Q_{\alpha}(0)}{\bar{B}(P)}
  \IEyn\IEysn\label{eq:17a}\\
  &=&\frao{4}\mel{0}{\bar{q}(z)Q(0)}{\bar{B}}I_{\alpha\beta}
  +\frao{4}\mel{0}{\bar{q}(z)\gma^5Q(0)}
  {\bar{B}}(\gma^5)_{\alpha\beta}\IEnn\\
  &&+\frao{8}\mel{0}{\bar{q}(z)\sigma^{\mu\nu}\gma^5Q(0)}
  {\bar{B}}(\sigma_{\mu\nu}\gma^5)_{\alpha\beta}
  +\frao{4}\mel{0}{\bar{q}(z)\gma^{\mu}Q(0)}
  {\bar{B}}(\gma_{\mu})_{\alpha\beta}\IEnn\\
  &&-\frao{4}\mel{0}{\bar{q}(z)\gma^{\mu}\gma^5Q(0)}
  {\bar{B}}(\gma_{\mu}\gma^5)_{\alpha\beta},
  \IEysn\label{eq:17b}
\end{IEEEeqnarray}
where $\sigma^{\mu\nu}=\frac{i}{2}\qty[\gma^{\mu},\gma^{\nu}]$, and
$[z,0]$ stands for the path-ordered exponential, which is called
Wilson line that connects the point $0$ and $z$. The definition of
Wilson line is
\begin{IEEEeqnarray}{cx}
  [z,0]\equiv\mathrm{P}\mathrm{exp}
  \qty(i\!\int_0^z\dd{x^{\mu}}\!A_{\mu}(x)).\label{eq:18}
\end{IEEEeqnarray}

According to discrete symmetries of $C,P$ and $T$ , the matrix
elements in the right-hand side of Eq.\eqref{eq:17b} are related to
four DAs $\tilde{\phi}_i$ ($i=P,T,A1,A2$ ) as defined in
Ref.\cite{Grozin:1996pq}
\begin{IEEEeqnarray}{rCl}
  \mel{0}{\bar{q}(z)Q(0)}{\bar{B}}&=&0,
  \IEyn\IEysn\label{eq:19a}\\
  \mel{0}{\bar{q}(z)\gma^5Q(0)}{\bar{B}}&=&-if_Bm_B\tilde{\phi}_P,
  \IEysn\label{eq:19b}\\
  \mel{0}{\bar{q}(z)\sigma^{\mu\nu}\gma^5Q(0)}{\bar{B}}&=&
  -if_B\tilde{\phi}_T\qty(P^{\mu}z^{\nu}-P^{\nu}z^{\mu}),
  \IEysn\label{eq:19c}\\
  \mel{0}{\bar{q}(z)\gma^{\mu}Q(0)}{\bar{B}}&=&0,
  \IEysn\label{eq:19d}\\
  \mel{0}{\bar{q}(z)\gma^{\mu}\gma^5Q(0)}{\bar{B}}&=&
  f_B\qty(i\tilde{\phi}_{A1}P^{\mu}-m_B\tilde{\phi}_{A2}z^{\mu}),
  \IEysn\label{eq:19e}
\end{IEEEeqnarray}
where the DAs $\tilde{\phi}_i$ are functions of the coordinate $z$. In
our scenario, we calculate these five matrix elements in the $B$-meson
rest frame by using the $B$ meson state defined in Eq.(\ref{eq:2}). We
confirmed that the matrix elements in Eq.\eqref{eq:19a} and
Eq.\eqref{eq:19c} are indeed zero
\begin{IEEEeqnarray}{cx}
  \mel{0}{\bar{q}(z)Q(0)}{\bar{B}}=
  \mel{0}{\bar{q}(z)\gma^{\mu}Q(0)}{\bar{B}}=0.\label{eq:20}
\end{IEEEeqnarray}
For the pseudoscalar DA in Eq.\eqref{eq:19b}, we obtain
\begin{IEEEeqnarray}{cx}
  \tilde{\phi}_P(z)=N_B\int\!\dd[3]{k}\varPsi_0(\vv{k})
  \frac{-\qty[(E_q+m_q)(E_Q+m_Q)+|\vv{k}|^2]}
  {\sqrt{E_qE_Q(E_q+m_q)(E_Q+m_Q)}}e^{-ik_q\cdot z},\label{eq:21}
\end{IEEEeqnarray}
where $k_q^{\mu}=(E_q,\vv{k})$ is the four-momentum of the light quark
in the meson rest frame, and
\begin{equation}
 N_B\equiv\frac{i}{f_B}\sqrt{\frac{3}{(2\pi)^3m_B}}.
\end{equation}
It should be understood that the wave function $\varPsi_0(\vv{k})$ may
have an arbitrary phase which can be adjusted to obtain a positive
real decay constant according to the definition in Eq.\eqref{eq:1}.

For the other DAs in Eqs.\eqref{eq:19c} and \eqref{eq:19e} (the
detailed derivation can be found in Appendix A), we introduce two
functions $A_T$ and $A$ at first,
\begin{IEEEeqnarray}{rCl}
  A_T(k^1,k^2,k^3)&\equiv&\varPsi_0(\vv{k})
  \frac{E_Q+m_Q+E_q+m_q}{\sqrt{E_qE_Q(E_q+m_q)(E_Q+m_Q)}},
  \IEyn\IEysn\label{eq:22a}\\
  A(k^1,k^2,k^3)&\equiv&\varPsi_0(\vv{k})
  \frac{E_Q+m_Q-E_q-m_q}{\sqrt{E_qE_Q(E_q+m_q)(E_Q+m_Q)}},
  \IEysn\label{eq:22b}
\end{IEEEeqnarray}
where $k^1,k^2,k^3$ are the components of the light quark momentum
$\vv{k}$ , i.e., $\vv{k}=(k^1,k^2,k^3)$. Then we obtain the DAs as
\begin{IEEEeqnarray}{rCl}
  \tilde{\phi}_T(z)&=&N_B
  \int\!\dd[3]{k}\qty[\frao{3}\sum_i\int_0^{k^i}\!\!
  A_T(\eta,\dots)\eta\dd{\eta}]e^{-ik_q\cdot z},
  \IEyn\IEysn\label{eq:23a}\\
  \tilde{\phi}_{A2}(z)&=&N_B
  \int\!\dd[3]{k}\qty[\frao{3}\sum_i\int_0^{k^i}\!\!
  A(\eta,\dots)\eta\dd{\eta}]e^{-ik_q\cdot z},
  \IEysn\label{eq:23b}\\
  \tilde{\phi}_{A1}(z)&=&-N_B
  \int\!\dd[3]{k}e^{-ik_q\cdot z}\IEnn\\
  &&\cdot \qty[\varPsi_0(\vv{k})\frac{(E_q+m_q)(E_Q+m_Q)-|\vv{k}|^2}
  {\sqrt{E_qE_Q(E_q+m_q)(E_Q+m_Q)}}+E_qA(k^1,k^2,k^3)].
  \IEysn\label{eq:23c}
\end{IEEEeqnarray}
For the details of the summation in the square parentheses containing
the ellipsis, see Eq.\eqref{eq:A3}.

Now, with Eqs.(\ref{eq:19a})$\sim$(\ref{eq:19e}), the matrix element
for $B$-meson in Eq.(\ref{eq:17a}) can be rewritten as
\begin{IEEEeqnarray}{rl}
  \tilde{\Phi}_{\alpha\beta}(z)=\frac{-if_B}{4}\Big\{
  \Big[m_B\tilde{\phi}_P&+\frao{2}\tilde{\phi}_{T}
  \qty(P^{\mu}z^{\nu}-P^{\nu}z^{\mu})
  \sigma_{\mu\nu}\IEnn\\
  &+\qty(\tilde{\phi}_{A1}P^{\mu}+im_B\tilde{\phi}_{A2}z^{\mu})
  \gma_{\mu}\Big]\gma^5\Big\}_{\alpha\beta},\label{eq:24}
\end{IEEEeqnarray}
where the DAs are given in Eqs.\eqref{eq:21} and
\eqref{eq:23a}--\eqref{eq:23c}.

In order to obtain the expressions of the DAs in momentum space, we
make use of the amplitude of a decay process which can be expressed as
a convolution \cite{Beneke:2000wa}
\begin{IEEEeqnarray}{cx}
  F=\int\!\dd[4]{z}\tilde{\Phi}_{\alpha\beta}(z)
  \tilde{T}_{\beta\alpha}(z).\label{eq:25}
\end{IEEEeqnarray}
Substituting Eq.\eqref{eq:24} into Eq.\eqref{eq:25} and with a few
steps of calculation (see Appendix B for details), we obtain
\begin{IEEEeqnarray}{rCl}
  \Phi_{\alpha\beta}(l)&=\Bigg\{\frac{-if_Bm_B}{4}\Bigg[\phi_P(l)
  &+\frac{i}{2}\phi_{T}(l)\sigma_{\mu\nu}
  \qty(v^{\mu}\pdv{l_{\nu}}-v^{\nu}\pdv{l_{\mu}})\IEnn\\
  &&+\qty(\phi_{A1}(l)\sld{v}-\phi_{A2}(l)\gma_{\mu}\pdv{l_{\mu}})
  \Bigg]\gma^5\Bigg\}_{\alpha\beta}\label{eq:26}
\end{IEEEeqnarray}
and
\begin{IEEEeqnarray}{cx}
  F=\int\!\dd[3]{l}\Phi_{\alpha\beta}(l)
  \eval{T_{\beta\alpha}(l)}_{l^2=m_q^2}.\label{eq:27}
\end{IEEEeqnarray}
It is understood that the derivative $\pdv{l_{\mu,\nu}}$ in
Eq.\eqref{eq:26} (which is called the momentum space projector
\cite{Beneke:2000wa,Wei:2002iu}) acts on the hard-scattering kernel
$T_{\beta\alpha}(l)$ before $l=k_q$ is taken. For the DAs in the
momentum space, we obtain
\begin{IEEEeqnarray}{rCl}
  \phi_P(k_q^{\mu})&=&-N_B\frac{(E_q+m_q)(E_Q+m_Q)+|\vv{k}|^2}
  {\sqrt{E_qE_Q(E_q+m_q)(E_Q+m_Q)}}\varPsi_0(\vv{k}),
  \IEyn\IEysn\label{eq:28a}\\
  \phi_T(k_q^{\mu})&=&\frac{N_B}{3}\sum_i\int_0^{k^i}\!\!A_T(\eta,\dots)
  \eta\dd{\eta},\IEysn\label{eq:28b}\\
  \phi_{A2}(k_q^{\mu})&=&\frac{N_B}{3}\sum_i\int_0^{k^i}\!\!A(\eta,\dots)
  \eta\dd{\eta},\IEysn\label{eq:28c}\\
  \phi_{A1}(k_q^{\mu})&=&-N_B\qty[\varPsi_0(\vv{k})
  \frac{(E_q+m_q)(E_Q+m_Q)-|\vv{k}|^2}{\sqrt{E_qE_Q(E_q+m_q)(E_Q+m_Q)}}
  +E_qA(k^1,k^2,k^3)],\IEysn\label{eq:28d}
\end{IEEEeqnarray}
In general, these DAs play an important role in the study of the
$B$-meson decays \cite{Lee:2005gza}. Thus it is necessary and useful
to give an numerical illustration of them.

\begin{figure}[tbp]
  \centering
  \includegraphics[width=\textwidth]{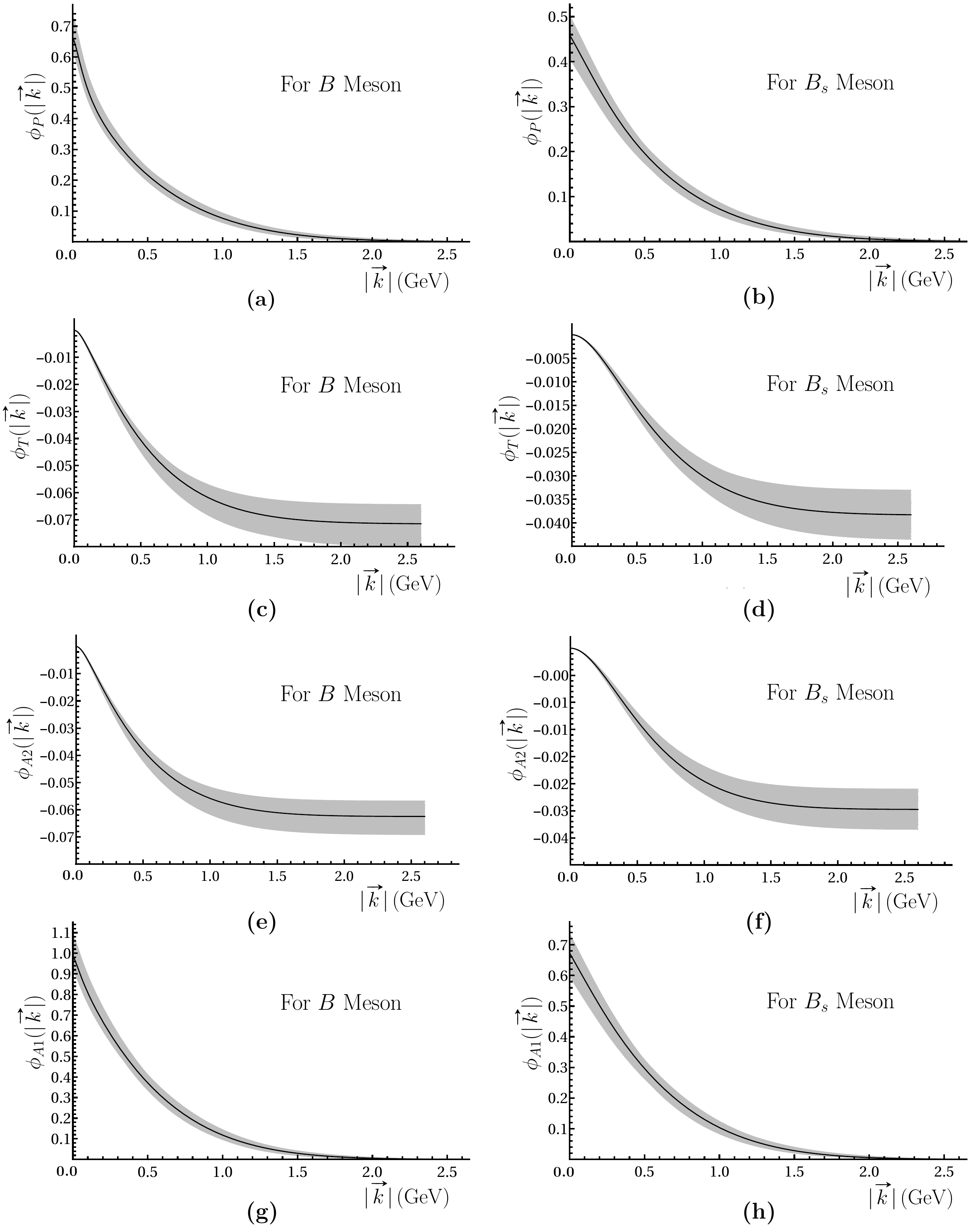}
  \caption{\label{fig:3}Distribution amplitudes as functions of
    $|k^3|$, where the grey bands are uncertainties caused by the wave
    function.}
\end{figure}

For simplicity, we take $\vv{k}=(0,0,k^3)$ and the DAs as functions of
$|k^3|$ are shown in Fig.\ref{fig:3}. The grey bands are the possible
uncertainties caused by the uncertainty of the wave function. In the
heavy-quark limit, one can obtain that one of the axial-vector DA
$\phi_{A2}$ is equal to the axial-tensor DA $\phi_T$
\cite{Grozin:1996pq}. For our results, as shown in Figs.\ref{fig:3}
(c) and (e), (d) and (f), these two DAs are indeed very close, which
indicate that our scenario is reasonable and their difference reflects
the influence of the finite heavy-quark mass.

One can also see that the figures for $B$ and $B_s$ mesons are very
similar, but in detail, for the same values of $|k^3|$, the absolute
values of the DAs of $B$ meson are always a bit larger than that of
$B_s$ meson. This is consistent with the fact that the DAs are
inversely proportional to the square root of the decay constants and
masses.

In addition, the light-cone coordinate is widely used in the study of
the DAs , for example, the works in
Refs.\cite{Beneke:2000wa,Lee:2005gza,Grozin:2005iz,Wei:2002iu,Bell:2008er,Bell:2013tfa,Feldmann:2014ika}
and references therein, where the DAs depend on a single variable
$k_+$ or $k_-$, which are the light-cone projections of the momentum
of the light antiquark in the rest frame of the meson . The
definitions of the light-cone projections of the momentum of the light
antiquark are
\begin{equation}
k_{\pm}=\frac{E_q\pm k^3}{\sqrt{2}},\quad{}
k_{\perp}^{\mu}=(0,k^1,k^2,0)
\end{equation}
Performing the integration over the transverse momentum $k_{\perp}$,
we can obtain the light-cone distribution amplitudes (LCDAs) in our
scenario. Usually, the $k_{\perp}$-integral is restricted by a scale
$\mu$, i.e., $|k_{\perp}|<\mu$ \cite{Ball:1998je,Beneke:2000ry}. In
our model, the wave function is spherically symmetric with respect to
$k^1$, $k^2$, and $k^3$. The integral region of the $k_{\perp}$ has an
upper limit, which is determined by Eqs.\eqref{eq:9a}--\eqref{eq:9c}.
The upper limits are shown clearly by the cut lines in
Fig.\ref{fig:1}.

\begin{figure}[tbp]
  \centering
  \includegraphics[width=\textwidth]{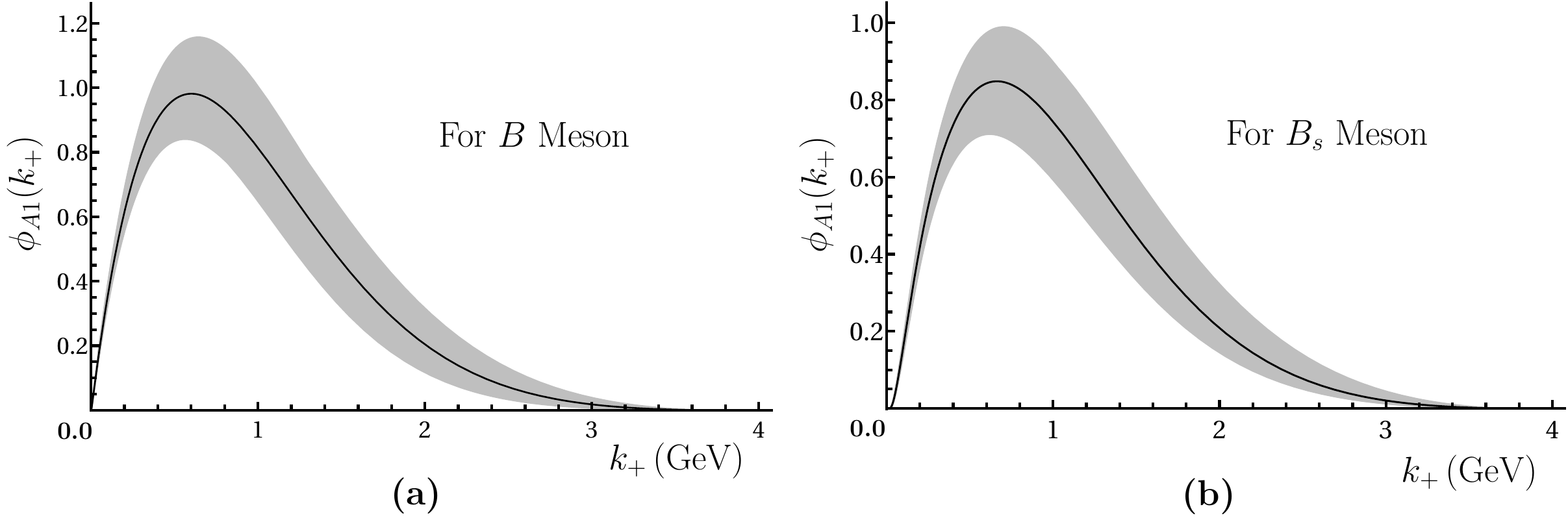}
  \caption{\label{fig:4}Distribution amplitudes as functions of $k_+$,
    where the grey bands are uncertainties caused by the wave
    function.}
\end{figure}

The distribution amplitude $\phi_{A1}$ as a function of $k_+$ is shown
in Fig.\ref{fig:4}. $\phi_{A1}$ is relevant to the LCDA $\phi_B^+$ in
the heavy quark limit, which is generally used in the study of $B$
decays . Our results are consistent with the general analysis given in
Ref.\cite{Lee:2005gza}.

Next we try to give a compact form of the matrix element
$\tilde{\Phi}_{\alpha\beta}(z)=\mel{0}
{\bar{q}_{\beta}(z)[z,0]Q_{\alpha}(0)}{\bar{B}(P)}$. Substituting
Eq.\eqref{eq:21} and Eqs.\eqref{eq:23a}--\eqref{eq:23c} into
Eq.\eqref{eq:17b} and after a few steps of simplification, we obtain
\begin{IEEEeqnarray}{rcl}
  \tilde{\Phi}_{\alpha\beta} (z)&=&\frac{-1}{4}
  \sqrt{\frac{3m_B}{(2\pi)^3}}\!\int\!\dd[3]{k}
  \frac{\varPsi_0(\vv{k})e^{-i k_q\cdot z}}
  {\sqrt{E_qE_Q(E_q+m_q)(E_Q+m_Q)}}
  \qty{\mqty(b\\c)\mqty(c&a)}_{\alpha\beta}\text{(\,D.R.)}
  \IEyn\IEysn\label{eq:29a}\\
  &=&\frac{-1}{4}\sqrt{\frac{3m_B}{(2\pi)^3}}\int\dd[3]{k}
  \cdot{}\IEnn\\
  &&\quad\frac{\varPsi_0(\vv{k}) e^{-ik_q\cdot z}}{\sqrt{E_qE_Q(E_q+m_q)(E_Q+m_Q)}}
  \qty{\mqty(b-c\\b+c)\mqty(c-a&c+a)}_{\alpha\beta}\text{(\,W.R.)}.
  \IEysn\label{eq:29b}
\end{IEEEeqnarray}
where $a$, $b$, and $c$ are three $2\!\times{}\!2$ matrices, which are
defined as
\begin{equation*}
  a=(E_q+m_q)I_{2\times2}, \;
  b=(E_Q+m_Q)I_{2\times 2},\;
  c=\vv{k}\cdot\vv{\sigma}
\end{equation*}
and $\vv{\sigma}$ is the Pauli matrix. These two expressions in
Eqs.\eqref{eq:29a} and \eqref{eq:29b} are derived with different
representations of the gamma matrix $\gma^{\mu}$. The label D.R.
denotes Dirac representation, and W.R. Weyl representation.

For simplicity, we define
\begin{IEEEeqnarray}{cx}
  K(\vv{k})\equiv\frac{-N_B\varPsi_0(\vv{k})}
  {\sqrt{E_qE_Q(E_q+m_q)(E_Q+m_Q)}}.\label{eq:30}
\end{IEEEeqnarray}
Then the convolution formula of Eq.\eqref{eq:27} can be rewritten as
\begin{IEEEeqnarray}{cx}
  F=\int\!\dd[3]{k}\frac{-if_Bm_B}{4}K(\vv{k})
  \qty{\mqty(b\\c)\mqty(c&a)}_{\alpha\beta}
  \eval{T_{\beta\alpha}(k_q)}_{k_q^2=m_q^2}\label{eq:31}
\end{IEEEeqnarray}
where the spinor matrices are given in Dirac representation (D.R.).

Next we introduce two light-like vectors $n_{\pm}^{\mu}=(1,0,0,\mp 1)$
and define $\sld{n}_+\equiv{}n_+^{\mu}\gma_\mu=\left(\smqty{1 &
    \sigma{}^3 \\ -\sigma{}^3 & -1}\right)$, $\sld{n}_-\equiv
n_-^{\mu}\gma_{\mu}=\left(\smqty{1 & -\sigma^3 \\ \sigma^3 &
    -1}\right)$. With these two vectors $n_{\pm}^{\mu}$, the matrix
element $\Phi_{\alpha\beta}(k_q^\mu)$ can be expressed in another form
\begin{IEEEeqnarray}{rCl}
  \Phi_{\alpha\beta}(k_q^{\mu})&=&\frac{-if_Bm_B}{4}K(\vv{k})
  \qty{\mqty(b\\c)\mqty(c&a)}_{\alpha\beta}\IEnn\\
  &=&\frac{-if_Bm_B}{4}K(\vv{k})\IEnn\\
  &&\cdot\Bigg\{(E_Q+m_Q)\frac{1+\sld{v}}{2}\Bigg[\qty(\frac{k_+}{\sqrt{2}}
  +\frac{m_q}{2})\sld{n}_++\qty(\frac{k_-}{\sqrt{2}}
  +\frac{m_q}{2})\sld{n}_--k_{\perp}^{\mu}\gma_{\mu}
  \Bigg]\gma^5\IEnn\\
  &&-(E_q+m_q)\frac{1-\sld{v}}{2}\Bigg[
  \qty(\frac{k_+}{\sqrt{2}}-\frac{m_q}{2})\sld{n}_+
  +\qty(\frac{k_-}{\sqrt{2}}-\frac{m_q}{2})\sld{n}_-
  -k_{\perp}^{\mu}\gma_{\mu}\Bigg]\gma^5
  \Bigg\}_{\alpha\beta}.\label{eq:32}
\end{IEEEeqnarray}
Compared with the commonly used results (for instance, see Eq.(109) in
Ref.\cite{Beneke:2000wa} and Eq.(2.48) in Ref.\cite{Bell:2013tfa}),
this new form includes the whole spinor structure of the momentum
projector. The part containing $\frac{1+\sld{v}}{2}$ is proportional
to the heavy quark's mass and is the only term in the heavy quark
limit. Since when the heavy-quark mass $m_Q$ goes infinity, the
contribution of other part in Eq.\eqref{eq:32} will be relatively very small and
can be ignored. Therefore, as we take the finite heavy-quark mass, the
part with $(E_q+m_q)$ will give extra contribution and may be an
important correction in the study of $B$-meson decays.
\section{QCD One-Loop Corrections to Leptonic Decays of $B$-Meson}
\label{sec:qcdf}

In Sec.\ref{sec:dc}, we study the leptonic decays of $B$ meson at tree
level. In this section, we extend this study by including QCD one-loop
corrections. When considering one-loop corrections in QCD, if one
naively calculate the loop diagrams, one will encounter not only
ultraviolet divergence, but also infrared divergence. Factorization
method can be applied to obtain the infrared-safe amplitude at the
quark level. To obtain the infrared-safe transition amplitude at quark
level, let us consider the free quark state
$\ket{\bar{u}^r(k)b^s(p-k)}$ as the initial state at first.
Factorization means that the matrix element of a physics transition
process $F^\mu$ can be expressed as the convolution of the wave
function of the initial state and the hard transition amplitude $T$
\begin{equation}
  F^\mu=\Phi{}\otimes{}T
\end{equation}
where the circle-time $\otimes$ denotes the convolution in
Eq.\eqref{eq:25}, and $\mu$ denotes the Lorentz index that may appear
in the physical transition matrix element. All the infrared
contributions are absorbed into the wave function $\Phi$, while the
hard amplitude $T$ is infrared safe.

In perturbation theory, the matrix element $F^{\mu}$, which relevant
to the quark transition process, the wave function $\Phi$ and the
hard-scattering kernel $T$ can all be expanded by the power of
$\alpha_s$. Therefore the factorization formula takes the form
\cite{DescotesGenon:2002mw}
\begin{IEEEeqnarray}{rCl}
  F^{\mu}&=&F^{(0)\mu}+F^{(1)\mu}+\cdots=\Phi\otimes T\IEnn\\
  &=&\qty[\Phi^{(0)}\otimes T^{(0)}]+\qty[\Phi^{(0)}\otimes
  T^{(1)}+\Phi^{(1)}\otimes T^{(0)}]+\cdots,\label{eq:33}
\end{IEEEeqnarray}
where the superscripts $(n)$'s indicate the perturbation levels. After
calculating both the matrix element $F^{(1)\mu}$ and the wave function
$\Phi^{(1)}$ at one-loop order, one can extract the hard amplitude
$T^{(1)}$ by using Eq.\eqref{eq:33}, that is
\begin{equation}
  \Phi^{(0)}\otimes{}T^{(1)}=F^{(1)\mu}-\Phi^{(1)}\otimes T^{(0)} \label{eq:34}
\end{equation}
At one-loop level, both the matrix element $F^{(1)\mu}$ and the wave
function $\Phi^{(1)}$ are infrared divergent. Through the subtraction
in the right-hand side of Eq.\eqref{eq:34}, the infrared divergence
can be cancelled . Then the hard amplitude $T^{(1)}$ we get through
Eq.\eqref{eq:34} is infrared safe.

\begin{figure}[tbp]
  \centering
  \includegraphics[width=0.4\textwidth]{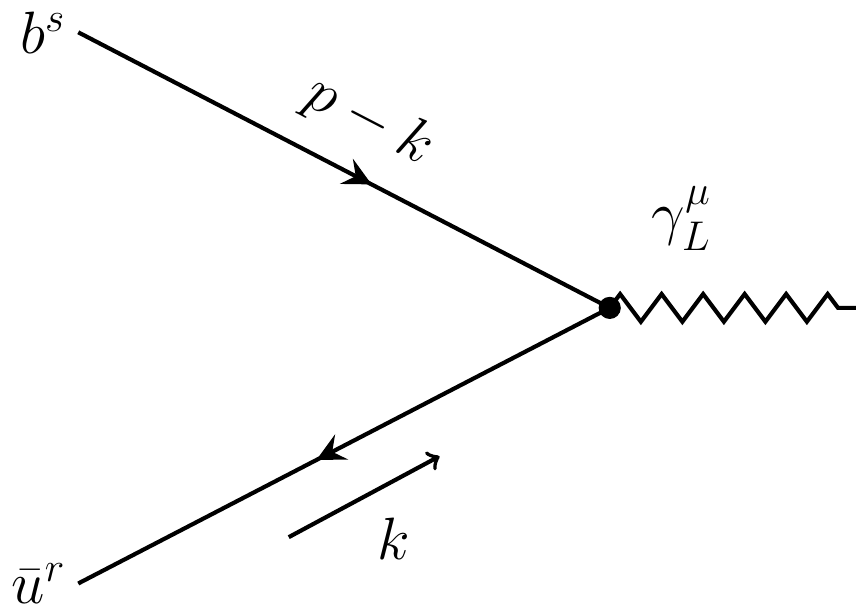}
  \caption{\label{fig:5}Factorization at tree level}
\end{figure}

At tree level, the factorization can be achieved straightforwardly and
we show the results briefly at first. The matrix element $F^\mu$ at
tree level, as shown in Fig. \ref{fig:5}, can be obtained as
\begin{IEEEeqnarray}{rCl}
  F_{b\bar{u}}^{(0)\mu}&=&\mel{0}{\bar{u}\gma_L^{\mu}b}
  {\bar{u}^r(k)b^s(p-k)}\IEnn\\
  &=&\frao{(2\pi)^3}\sqrt{\frac{m_um_b}{k^0(p-k)^0}}
  \bar{v}^r(k)\gma_L^{\mu}u^s(p-k)\label{eq:35}
\end{IEEEeqnarray}
where the coefficient
$\frao{(2\pi)^3}\sqrt{\frac{m_um_b}{k^0(p-k)^0}}$ is from our
convention, and $\bar{v}$ and $u$ are the spinors of the quarks
$\bar{u}$ and $b$, respectively. The superscripts $r$ and $s$ are the
spin labels.

The wave function of the free quark state $\ket{\bar{u}^r(k)b^s(p-k)}$
at tree level is
\begin{IEEEeqnarray}{rCl}
  \Phi_{\alpha\beta}^{(0)b\bar{u}}(\tilde{k})&=&
  \int\!\dd[4]{z}e^{i\tilde{k}\cdot z}
  \mel{0}{\bar{u}_{\beta}(z)[z,0]b_{\alpha}(0)}{\bar{u}^r(k)b^s(p-k)}
  \IEnn\\
  &=&\frao{(2\pi)^3}\sqrt{\frac{m_um_b}{k^0(p-k)^0}}
  (2\pi)^4\dlt{4}{\tilde{k}-k}\bar{v}^r_{\beta}(k)u^s_{\alpha}(p-k)
  .\label{eq:36}
\end{IEEEeqnarray}

Matching the matrix element in Eq.\eqref{eq:35} and the wave function
in Eq.\eqref{eq:36} into the factorization formula
\begin{IEEEeqnarray}{rCl}
  F^{(0)\mu}_{b\bar{u}}&=&\int\!\frac{\dd[4]{\tilde{k}}}{(2\pi)^4}
  \Phi_{\alpha\beta}^{(0)b\bar{u}}(\tilde{k})T_{\beta\alpha}^{(0)}(\tilde{k})
  \IEnn\\
  &=&\frao{(2\pi)^3}\sqrt{\frac{m_um_b}{k^0(p-k)^0}}\bar{v}_{\beta}^r(k)
  T_{\beta\alpha}^{(0)}(k)u_{\alpha}^s(p-k),\label{eq:37}
\end{IEEEeqnarray}
we can obtain the hard-scattering kernel at tree level
\begin{IEEEeqnarray}{cx}
  T_{\beta\alpha}^{(0)}(k)=\qty(\gma_L^{\mu})_{\beta\alpha}.\label{eq:38}
\end{IEEEeqnarray}
This tree-level result is independent of the quark momentum $k$. It
plays an important role in the calculation of the hard amplitude at
one-loop level.

\begin{figure}[bp]
  \centering
  \includegraphics[width=1.0\textwidth]{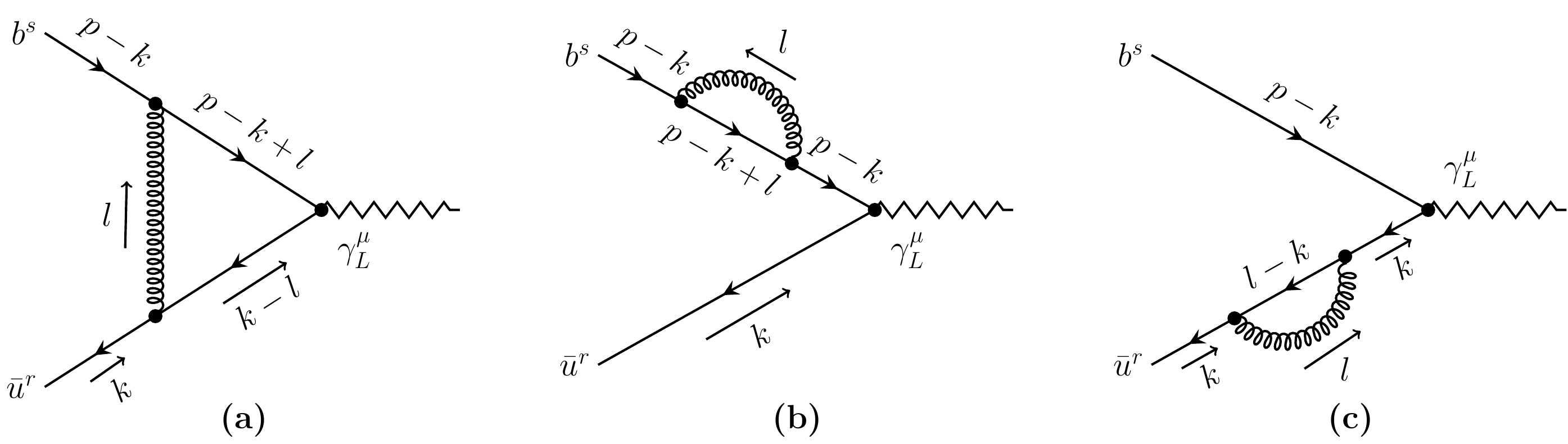}
  \caption{\label{fig:6}Feynman diagrams at one-loop level for
    $F$.}
\end{figure}

Next we shall establish the factorization at one-loop level. The
Feynman diagram for the matrix element $F^{(1)\mu}$ at one-loop level
is shown as Fig. \ref{fig:6}(a). The renormalization factor
$\sqrt{Z_2^{\bar{u}} Z_2^b}$ must appear in the contribution of Fig.
\ref{fig:6}(a) due to the renormalization of the external quark
fields, where $\sqrt{Z_2^{\bar{u}}}$ and $\sqrt{Z_2^b}$ are the
renormalization constants of the external quark fields $\bar{u}$ and
$b$, respectively. Since the factor $\sqrt{Z_2^{b}}$ and
$\sqrt{Z_2^{\bar{u}}}$ correspond to the self-energy diagrams of the
external quark $b$ and $\bar{u}$, the factor
$\sqrt{Z_2^{\bar{u}}Z_2^b}$ can be represented by the contributions of
Fig. \ref{fig:6}(b) and (c) .

The contribution of Fig. \ref{fig:6} (a) is
\begin{IEEEeqnarray}{rCl}
  F_V^{(1)\mu}&=&\frao{(2\pi)^3}\sqrt{\frac{m_um_b}{k^0(p-k)^0}}
  \qty(-ig_s^2)C_F\bar{v}^r(k)\IEnn\\
  &&\cdot\int\!\frac{\dd[4]{l}}{(2\pi)^4}
  \gma^{\rho}\frao{m_u-\qty(\sld{l}-\sld{k})}\gma_L^{\mu}
  \frao{m_b-\qty(\sld{p}-\sld{k}+\sld{l})}\gma_{\rho}\frao{l^2}
   u^s(p-k), \label{eq:39}
\end{IEEEeqnarray}
where $g_s$ is the strong coupling constant, and all the momenta of
quarks and gluon are labelled in Fig. \ref{fig:6} (a). The explicit
result after the loop integration is given in Appendix C.

The contributions of Fig. \ref{fig:6} (b) and (c)  are
\begin{IEEEeqnarray}{cx}
  F_{bR}^{(1)\mu}=\frao{2}\qty(Z_2^b-1)F_{b\bar{u}}^{(0)\mu}\qc
  F_{\bar{u}R}^{(1)\mu}=\frao{2}\qty(Z_2^{\bar{u}}-1)F_{b\bar{u}}^{(0)\mu}.
  \label{eq:40}
\end{IEEEeqnarray}
The renormalization constants (the explicit expressions are listed in
Appendix C) are defined in terms of the one-particle
irreducible (1PI) diagrams $\Sigma$ by
\begin{IEEEeqnarray}{cx}
  Z_2^{b,\bar{u}}=1+i\eval{\dv{\Sigma}{\sld{p}}}_{\sld{p}=m}.\label{eq:41}
\end{IEEEeqnarray}

\begin{figure}
  \centering
  \includegraphics[width=1.0\textwidth]{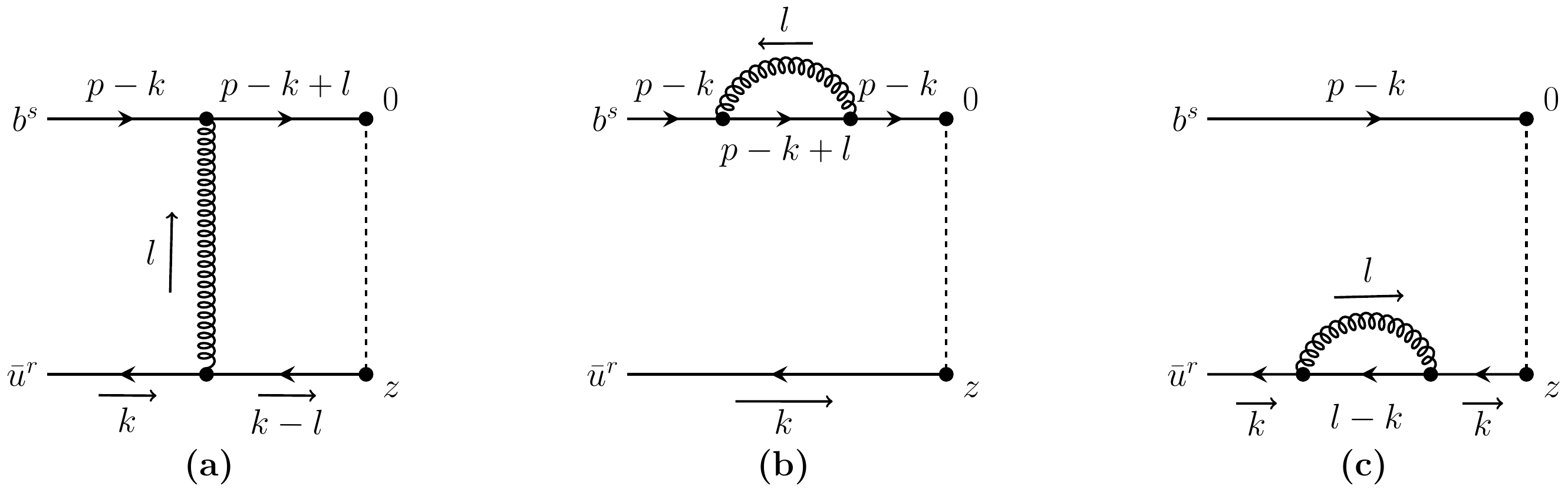}
  \caption{\label{fig:7}Feynman diagrams at one-loop level for
     WF (1).}
\end{figure}

\begin{figure}[tbp]
  \centering
  \includegraphics[width=1.0\textwidth]{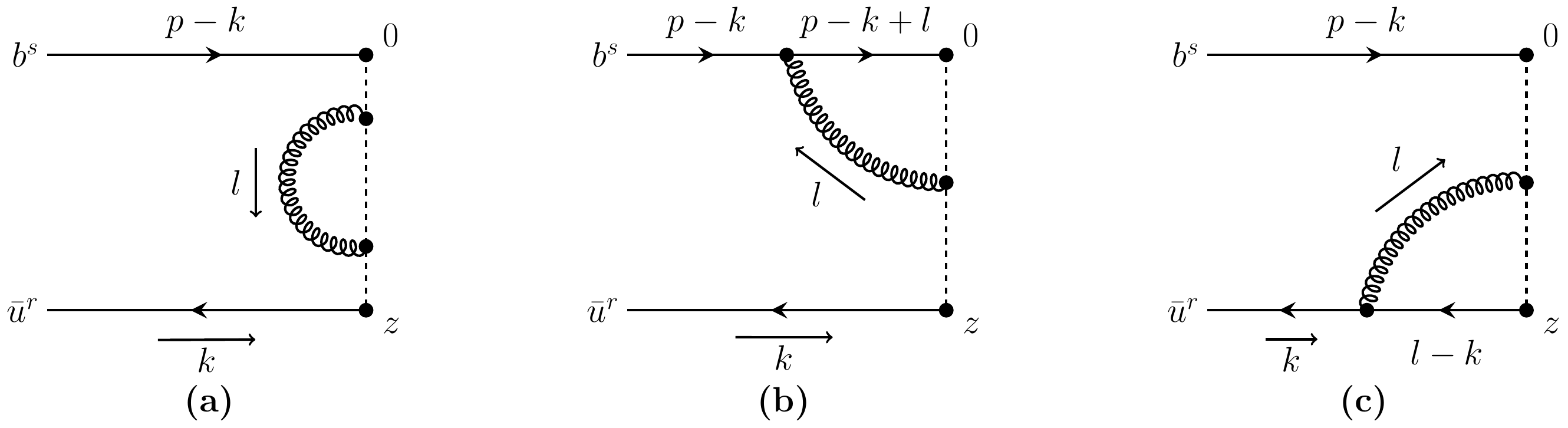}
  \caption{\label{fig:8}Feynman diagrams at one-loop level for
    WF (2).}
\end{figure}

The corrections for the wave functions at one-loop order contain 6
Feynman diagrams which have been divided into two groups. They are
shown in Figs. \ref{fig:7} and \ref{fig:8}. It will be shown later
that, when the contribution of the diagrams in Fig. \ref{fig:8} is
convoluted with the hard-transition kernel at tree level, the result
will be zero.

The contribution of the diagram Fig. \ref{fig:7}(a)  to the wave function is
\begin{IEEEeqnarray}{rCl}
  \Phi_{\alpha\beta}^{(1)V}(\tilde{k})&=&(2\pi)^4\dlt{4}{l-k+\tilde{k}}
  \frao{(2\pi)^3}\sqrt{\frac{m_um_b}{k^0(p-k)^0}}
  (-ig_s^2)C_F\int\!\frac{\dd[4]{l}}{(2\pi)^4}\cdot\IEnn\\
  &&\qty[\bar{v}^r(k)\gma^{\rho}\frao{m_u-\qty(\sld{l}-\sld{k})}]_{\beta}
  \qty[\frao{m_b-\qty(\sld{p}-\sld{k}+\sld{l})}\gma_{\rho}\frao{l^2}
  u^s(p-k)]_{\alpha}\label{eq:42}
\end{IEEEeqnarray}
and the contributions of the wave function renormalization of the
heavy quark field [Fig. \ref{fig:7}(b)] and the light quark field
[Fig. \ref{fig:7}(c)] are
\begin{IEEEeqnarray}{cx}
  \Phi_{\alpha\beta}^{(1)b}=\frao{2}\qty(Z_2^b-1)\Phi_{\alpha\beta}^{(0)b\bar{u}}
  \qc\Phi_{\alpha\beta}^{(1)\bar{u}}=
  \frao{2}\qty(Z_2^{\bar{u}}-1)\Phi_{\alpha\beta}^{(0)b\bar{u}}.\label{eq:43}
\end{IEEEeqnarray}
Then it is straightforward to obtain the results after the convolution
with the hard-scattering kernel at tree level
$T_{\beta\alpha}^{(0)}=\qty(\gma_L^{\mu})_{\beta\alpha}$ and we find
that
\begin{IEEEeqnarray}{rCl}
  F_V^{(1)\mu}&=&\Phi_{\alpha\beta}^{(1)V}\otimes T_{\beta\alpha}^{(0)}
  ,\IEyn\IEysn\label{eq:44a}\\
  F_{bR}^{(1)\mu}&=&\Phi_{\alpha\beta}^{(1)b}\otimes T_{\beta\alpha}^{(0)}
  ,\IEysn\label{eq:44b}\\
  F_{\bar{u}R}^{(1)\mu}&=&\Phi_{\alpha\beta}^{(1)\bar{u}}\otimes
  T_{\beta\alpha}^{(0)}.\IEysn\label{eq:44c}
\end{IEEEeqnarray}
It is noted that there are two scales in the above equations, i.e.,
the factorization scale $\mu_F$ in the wave functions
$\Phi_{\alpha\beta}^{(1)b,\bar{u}}$ and the renormalization scale
$\mu_R$ in the matrix element $F_{b,\bar{u}R}^{(1)\mu}$. Here we take
$\mu_F=\mu_R$.

At last, we turn to the contributions of Feynman diagrams in Fig.
\ref{fig:8}. The contribution of Fig. \ref{fig:8}(a) contains a gluon
propagator both the starting and ending points being on the
Wilson-line. In the light-cone approximation and working in the
Feynman gauge, this propagator vanishes \cite{DescotesGenon:2002mw}
since $z$ is a null vector on the light-cone ($z^2=0$). As for our
case, the result is still zero. First, we obtain
\begin{IEEEeqnarray}{rCl}
  \Phi_{\alpha\beta}^{(1)0a}(\tilde{k})&=&\frao{(2\pi)^3}
  \sqrt{\frac{m_um_b}{k^0(p-k)^0}}(ig_s^2)C_F
  \bar{v}^r(k)_{\beta}\int\!\frac{\dd[4]{l}}{(2\pi)^4}
  \int\!\dd[4]{z}e^{i\tilde{k}\cdot z}\cdot\IEnn\\
  &&2\int_0^1\!\dd{x}z^{\mu}\int_0^x\dd{y}z^{\nu}
  e^{-ik\cdot z}e^{ixl\cdot z}e^{iy(-l)\cdot z}\frac{g_{\mu\nu}}{l^2}
  u^s(p-k)_{\alpha}\IEyn\IEysn\label{eq:45a}\\
  &=&\frac{-i2g_s^2C_F}{(2\pi)^3}\sqrt{\frac{m_um_b}{k^0(p-k)^0}}
  \bar{v}^r(k)_{\beta}\int\!\frac{\dd[4]{l}}{(2\pi)^4}\int\!
  \dd[4]{z}\int_0^1\!\dd{x}\int_0^x\!\dd{y}\IEnn\\
  &&\frac{e^{i(xl-k-yl)\cdot z}}{l^2}
  \qty[\pdv{}{\tilde{k}_{\mu}}{\tilde{k}^{\mu}}
  e^{i\tilde{k}\cdot z}]u^s(p-k)_{\alpha}.\IEysn\label{eq:45b}
\end{IEEEeqnarray}
In  Eq.\eqref{eq:45a} we make the substitution
$x^{\mu}=xz^{\mu}$ and $y^{\nu}=yz^{\nu}$ in the Wilson-line.

Next, we can substitute Eq.\eqref{eq:45b} into the convolution
formula, and perform the partial integration. By noting that the
hard-scattering kernel is a constant Dirac matrix, we can demonstrate
\begin{IEEEeqnarray}{rCl}
  \Phi_{\alpha\beta}^{(1)0a}\otimes T_{\beta\alpha}^{(0)}
  &=&\int\!\frac{\dd[4]{\tilde{k}}}{(2\pi)^4}
  \frac{i2g_s^2C_F}{(2\pi)^3}\sqrt{\frac{m_um_b}{k^0(p-k)^0}}
  \bar{v}^r(k)_{\beta}\int\!\frac{\dd[4]{l}}{(2\pi)^4}\int\!
  \dd[4]{z}\int_0^1\!\dd{x}\int_0^x\!\dd{y}\IEnn\\
  &&\frac{e^{i(xl-k-yl)\cdot z}}{l^2}e^{i\tilde{k}\cdot z}
  \qty[\pdv{}{\tilde{k}_{\mu}}{\tilde{k}^{\mu}}
  T_{\beta\alpha}^{(0)}]u^s(p-k)_{\alpha}\IEnn\\
  &=&0.\label{eq:46}
\end{IEEEeqnarray}

For the other two diagrams in Fig. \ref{fig:8}, the contribution of
Fig. \ref{fig:8}(b) is
\begin{IEEEeqnarray}{rCl}
  \Phi_{\alpha\beta}^{(1)0b}(\tilde{k})&=&\frac{i2g_s^2C_F}{(2\pi)^3}
  \sqrt{\frac{m_um_b}{k^0(p-k)^0}}
  \int\!\frac{\dd[4]{l}}{(2\pi)^4}\int\!\dd[4]{z}\int_0^1\!
  \dd{x}\frac{e^{i(xl-k)\cdot z}}{l^2}\IEnn\\
  &&\bar{v}^r(k)_{\beta}
  \qty[\pdv{}{\tilde{k}_{\rho}}e^{i\tilde{k}\cdot z}]
  \qty[\frao{m_b-(\sld{p}-\sld{k}+\sld{l})}
  \gma_{\rho}u^s(p-k)]_{\alpha}\label{eq:47}
\end{IEEEeqnarray}
Then the convolution is
\begin{IEEEeqnarray}{rCl}
  \Phi_{\alpha\beta}^{(1)0b}\otimes T_{\beta\alpha}^{(0)}
  &=&\int\!\frac{\dd[4]{\tilde{k}}}{(2\pi)^4}
  \frac{-i2g_s^2C_F}{(2\pi)^3}
  \sqrt{\frac{m_um_b}{k^0(p-k)^0}}
  \int\!\frac{\dd[4]{l}}{(2\pi)^4}\int\!\dd[4]{z}\int_0^1\!
  \dd{x}\frac{e^{i(xl-k)\cdot z}}{l^2}\IEnn\\
  &&\bar{v}^r(k)_{\beta}e^{i\tilde{k}\cdot z}
  \qty[\pdv{}{\tilde{k}_{\rho}}T_{\beta\alpha}^{(0)}]
  \qty[\frao{m_b-(\sld{p}-\sld{k}+\sld{l})}
  \gma_{\rho}u^s(p-k)]_{\alpha}\IEnn\\
  &=&0.\label{eq:48}
\end{IEEEeqnarray}
Similarly, we can obtain that the contribution of Fig. \ref{fig:8} (c)
is also zero.

Finally, combining Eqs.\eqref{eq:44a}--\eqref{eq:44c},
Eq.\eqref{eq:46} and Eq.\eqref{eq:48} together, we can demonstrate
that $F_{b\bar{u}}^{(1)\mu}=\Phi_{\alpha\beta}^{(1)b\bar{u}}\otimes
T_{\beta\alpha}^{(0)}$ and thus considering Eq.\eqref{eq:34}, the
total contribution to the hard-scattering kernel at one-loop level
$T_{\beta\alpha}^{(1)}$ is zero. Therefore the QCD one-loop
corrections to the hard amplitude of the leptonic decay of $B$ meson
are zero in the factorization scheme.

A brief remark about this result should be given here. The vanishment
of QCD one-loop correction to the hard decay amplitude of the pure
leptonic decay of $B$ meson does not mean that the naive calculation
of the QCD one-loop correction diagrams in Fig. \ref{fig:6} will
result in zero. The results in Eqs.\eqref{eq:39}, \eqref{eq:40} and
that given in Appendix C show that the contributions of these diagrams
are not zero. They include both hard and infrared singularities. The
infrared singularities come from the limit that the mass of the light
quark approaches zero and/or the momentum of the gluon vanishes, i.e.,
$m_q\to 0$ and $l\to 0$. It has been known that a conserved current
requires no renormalization because of gauge invariance \cite{boyer}.
Here the axial current $\bar{q}\gamma^\mu\gamma_5 b$ inducing the
leptonic decay of $B$ meson is partially conserved. Our calculation
shows that the axial current as a composite operator still does not
require renormalization. Only the external quark field renormalization
is needed. Although the naive contributions of the diagrams in Fig.
\ref{fig:6} are not zero, when QCD corrections to the wave function
are also considered up to one-loop order, the infrared singularities
and the hard contribution in the short-distance amplitude are
simultaneously subtracted by that in the wave function by using
Eq.\eqref{eq:34}. This implies that the infrared contribution in the
short-distance amplitude can be absorbed into the wave function, and
the hard terms are also absorbed and they will contribute to the
evolution of the wave function.

The factorization and the result that the hard amplitude receives no
QCD correction are proved up to one-loop order in this work. But we
expect that this result may hold up to all orders in QCD, because the
gluons are always restricted between the heavy quark and the light
antiquark lines for both the cases of the QCD corrections to the wave
function and that to the hard amplitude of the pure leptonic decay
process. Therefore the subtraction may happen up to all orders in
perturbative expansions. Then the formula that expresses the decay rate
of the leptonic decay in Eq.\eqref{eq:14} holds in all orders in
perturbation theory. QCD corrections can only change the theoretical
prediction to the decay constant.
\section{Discussion and Conclusion}
\label{sec:concl-or-disc}
Using the wave function that is obtained in the relativistic potential
model in our previous work \cite{Liu:2015lka}, where the hyperfine
interactions are included, the decay constants and pure leptonic
decays of $B$ meson are studied in this work. To keep the
four-momentum conservation between the quark-antiquark pair and the
meson, we use the ACCMM scenario
\cite{Altarelli:1982kh,Colangelo:1998eb} to treat the constituent
quarks, where the heavy quark is taken to be off-shell, while the
light antiquark is kept on shell. Compared with our earlier work
\cite{Yang:2011ie}, the difference is that the wave function used here
is obtained by considering the hyperfine interactions in the wave
equation, and the heavy quark is treated off-shell in the decay
process. The off-shellness of the heavy quark can be explained as
absorbing the effective effects of the gluon cloud around the heavy
quark. With such a treatment, the branching ratios of leptonic decays
of $B$ meson obtained in this work are consistent with experimental
data.

Based on the success of studying the leptonic decays of the $B$ meson,
we further obtain the distribution amplitudes for $B$ meson both in
coordinate and momentum space. The distribution amplitudes of $B$
meson are widely used in the study of $B$-meson decays. In addition,
we obtain another form of the nonlocal matrix element in
Eq.\eqref{eq:32}. Considering the success of the ACCMM scenario in
studying the leptonic decays, the heavy quark in the distribution
amplitude needs to be treated to be off-shell to maintain the momentum
and energy conservation. The new form of the nonlocal matrix element
obtained in this work, Eqs. (\ref{eq:29a}) or (\ref{eq:29b}) and
Eq.(\ref{eq:32}) should be useful in the study of the semileptonic
and nonleptonic $B$ decays, where the longitudinal and transverse
components are automatically included.

We finally study the QCD one-loop corrections within the frame work of
the factorization approach. We find that, after subtracting the
infrared divergence, the QCD one-loop corrections to the hard transition amplitude will be zero. This
implies that the infrared contributions in the hard amplitude can be absorbed into the wave
function and the hard terms originated from one-loop diagrams are also
absorbed by the wave function and they will contribute to the
evolution of the wave function. The formula expressing the leptonic
decay rate of $B$ meson in Eq. (\ref{eq:14}) is not affected by QCD
corrections.
\appendix
\section{DERIVATION OF THE DISTRIBUTION AMPLITUDES $\tilde{\phi}_T(z)$,
  $\tilde{\phi}_{A1}(z)$, AND $\tilde{\phi}_{A2}(z)$}
In this appendix, we give a brief derivation of the three distribution
amplitudes presented in Eqs.\eqref{eq:23a}-\eqref{eq:23c}. The direct
result about $\tilde{\phi}_T(z)$ in Eq.\eqref{eq:19c} is
\begin{IEEEeqnarray}{cx}
  \tilde{\phi}_T(z)z^i=iN_B\int\!\dd[3]{k}
  \varPsi_0(\vv{k})\frac{E_q+m_q+E_Q+m_Q}
  {\sqrt{E_qE_Q(E_q+m_q)(E_Q+m_Q)}}k^ie^{-ik_q\cdot z}\label{eq:A1}
\end{IEEEeqnarray}
where $k^i$ stands for any components of momentum $\vv{k}$ and
$N_B=\frac{i}{f_B}\sqrt{\frac{3}{(2\pi)^3m_B}}$. Note that
$z_ie^{-ik_q\cdot z}=i\pdv{k^i}e^{-ik_q\cdot z}$ and make use of
$A_T(k^1,k^2,k^3)$ defined in Eq.\eqref{eq:22a},
\begin{IEEEeqnarray}{Cl}
  &A_T(k^1,k^2,k^3)k^1=\pdv{k^1}\int_0^{k^1}\!A_T(\eta,k^2,k^3)
  \eta\dd{\eta}\IEyn\IEysn\label{eq:A2a}\\
  \Rightarrow&\tilde{\phi}_T(z)z^1=N_B\int\!\dd[3]{k}
  \!\int_0^{k^1}\!A_T(\eta,k^2,k^3)\eta\dd{\eta}
  (-z_1)e^{-ik_q\cdot z}\IEysn\label{eq:A2b}\\
  \Rightarrow&\tilde{\phi}_T(z)=N_B\int\!\dd[3]{k}
  \qty[\frao{3}\sum_i\int_0^{k^i}\!\!
  A_T(\eta,\dots)\eta\dd{\eta}]e^{-ik_q\cdot z}
  \IEysn\label{eq:A2c}
\end{IEEEeqnarray}
where Eq.\eqref{eq:A2b} is derived from Eq.\eqref{eq:A2a} by partial
integration. The summation in the square parentheses is short for the
following form
\begin{IEEEeqnarray}{rCl}
  \sum_i\int_0^{k^i}\!\!A_T(\eta,\dots)\eta\dd{\eta}&=&
  \int_0^{k^1}\!A_T(\eta,k^2,k^3)\eta\dd{\eta}
  +\int_0^{k^2}\!A_T(k^1,\eta,k^3)\eta\dd{\eta}\IEnn\\
  &&+\int_0^{k^3}\!A_T(k^1,k^2,\eta)\eta\dd{\eta}.\label{eq:A3}
\end{IEEEeqnarray}
The situation is similar for the derivation of $\tilde{\phi}_{A2}(z)$.

For the DA $\tilde{\phi}_{A1}(z)$, after substituting
Eq.\eqref{eq:23b} and Eq.\eqref{eq:22b} into the equation
Eq.\eqref{eq:19e}, we obtain
\begin{IEEEeqnarray}{rCl}
  \tilde{\phi}_{A1}(z)&=&
  -N_B\int\!\dd[3]{k}e^{-ik_q\cdot z}\IEnn\\
  &&\Bigg[\varPsi_0(\vv{k})
  \frac{(E_q+m_q)(E_Q+m_Q)-|\vv{k}|^2}
  {\sqrt{E_qE_Q(E_q+m_q)(E_Q+m_Q)}}
  +\frac{iz^0}{3}\sum_i\int_0^{k^i}\!
  A(\eta,\dots)\eta\dd{\eta}\Bigg]\label{eq:A4}
\end{IEEEeqnarray}
Using the same trick $iz^0e^{-ik_q\cdot z}=-\pdv{E_q}e^{-ik_q\cdot z}$
and partial integration and noting that in our scenario
$E_q^2-|\vv{k}|^2=m_q^2$, we get the final expression
\begin{IEEEeqnarray}{rCl}
  \tilde{\phi}_{A1}(z)&=&-N_B
  \int\!\dd[3]{k}e^{-ik_q\cdot z}\IEnn\\
  &&\qty[\varPsi_0(\vv{k})
  \frac{(E_q+m_q)(E_Q+m_Q)-|\vv{k}|^2}
  {\sqrt{E_qE_Q(E_q+m_q)(E_Q+m_Q)}}+E_qA(k^1,k^2,k^3)]
  \label{eq:A5}
\end{IEEEeqnarray}
\section{DERIVATION OF THE AMPLITUDE $F$ IN THE MOMENTUM SPACE}
In this appendix, we show explicitly how to derive Eq.\eqref{eq:27}
from Eq.\eqref{eq:25}. First, we perform the Fourier transformation
on the hard scattering kernel $\tilde{T}_{\beta\alpha}(z)$ and obtain
\begin{IEEEeqnarray}{cx}
  F=\!\int\!\dd[4]{z}\tilde{\Phi}_{\alpha\beta}(z)
  \!\int\!\frac{\dd[4]{l}}{(2\pi)^4}e^{il\cdot z}
  T_{\beta\alpha}(l)
  =\int\!\frac{\dd[4]{l}}{(2\pi)^4}\qty[\int\!\dd[4]{z}
  e^{i l\cdot z}\tilde{\Phi}_{\alpha\beta}(z)]
  T_{\beta\alpha}(l)\label{eq:B.1}
\end{IEEEeqnarray}
Performing Fourier transformation to the matrix element
$\tilde{\Phi}_{\alpha\beta}(z)$ in Eq.\eqref{eq:24}, and using
$z^{\mu}e^{il\cdot z}=-i\pdv{l_{\mu}}e^{il\cdot z}$ , we can obtain
\begin{IEEEeqnarray}{Lll}
  &\int\!\dd[4]{z}e^{il\cdot z}\tilde{\Phi}_{\alpha\beta}(z)
  =\frac{-if_B}{4}\!\int\!\dd[4]{z}\Bigg\{
  \Big[m_B\tilde{\phi}_P&+\frac{-i}{2}\tilde{\phi}_{T}
  \qty(P^{\mu}\pdv{l_{\nu}}-P^{\nu}\pdv{l_{\mu}})
  \sigma_{\mu\nu}\IEnn\\
  &&+\qty(\tilde{\phi}_{A1}P^{\mu}+m_B\tilde{\phi}_{A2}\pdv{l_{\mu}})
  \gma_{\mu}\Big]e^{il\cdot z}\cdot\gma^5
  \Bigg\}_{\alpha\beta}\label{eq:B.2}
\end{IEEEeqnarray}

Substituting Eq.(\ref{eq:B.2}) into Eq.(\ref{eq:B.1}), and making use
of partial integration, the derivative $\pdv{l_{\mu}}$ can be moved to
act on the hard scattering kernel $T_{\beta\alpha}(l)$. In addition,
we observe that in the four distribution amplitudes in
Eq.\eqref{eq:21} and Eqs.\eqref{eq:23a}--\eqref{eq:23c}, only the
exponential part $e^{-ik_q\cdot z}$ depends on the variable $z$.
Therefore the integration over $z$ can be easily worked out and the
result is a delta function $(2\pi)^4\dlt{4}{l-k_q}$.
\begin{IEEEeqnarray}{cx}
  F=\!\int\!\frac{\dd[4]{l}}{(2\pi)^4}(2\pi)^4\dlt{4}{l-k_q}
  \Big[\dots\Big]T_{\beta\alpha}(l)\label{eq:B.3}
\end{IEEEeqnarray}
After taking $P^{\mu}=m_Bv^{\mu}$ and $k_q^2=m_q^2$ [Eq.\eqref{eq:9b}]
into consideration, we obtain the final expression in Eq.\eqref{eq:27}.

\section{EXPLICIT EXPRESSIONS OF EQ.\eqref{eq:39} AND $Z_2^{b,\bar{u}}$}
We use the dimensional regularization for the ultraviolet divergence
and introduce a small mass $\lambda{}$ for gluons to regularize the
infrared divergence in Eq.\eqref{eq:39}. The naive dimensional regularization is adopted,
where $\gma^5$ anticommutes with all other gamma matrices.

The conventions and notations we use are
\begin{IEEEeqnarray*}{lll}
  \alpha_s=\frac{g_s^2}{4\pi}&\qquad{}C_F=\frac{N^2-1}{2N}&\qquad{}
  \sigma^{\mu\nu}=\frac{i}{2}[\gma^{\mu},\gma^{\nu}],\\
  \mathrm{P}_L=\frac{1-\gma^5}{2}&\qquad{}\mathrm{P}_R=\frac{1+\gma^5}{2}&\qquad{}
  \gma_{L}^{\mu}=\gma^{\mu}\mathrm{P}_L\qquad{}\gma_{R}^{\mu}=\gma^{\mu}\mathrm{P}_R.
\end{IEEEeqnarray*}

With the help of the program \emph{Package-X}
\cite{Patel:2015tea,Patel:2016fam}, the explicit result of Eq.\eqref{eq:39} is
\begin{IEEEeqnarray}{rCl}
  F_V^{(1)\mu}&=&\frao{(2\pi)^3}\sqrt{\frac{m_um_b}{k^0(p-k)^0}}
  \frac{\alpha_s C_F}{4\pi}\bar{v}^r(k)\cdot\text{\bfseries INT}\cdot{}u^s(p-k)
  \label{eq:C.1}
\end{IEEEeqnarray}
where
\begin{IEEEeqnarray}{rCl}
  \text{\bfseries INT}&=&\gma^{\mu}_L\cdot{}
  \qty[\frao{\varepsilon}-\gma_{\mathrm{E}}+\ln
  \frac{4\pi\mu^2}{m_um_b}+x\ln \frac{x+1}{x-1}\ln
  \frac{\lambda^2}{m_um_b}+F(x,x_1,x_2)]\IEnn\\
  &&+\gma^{\mu}_R\cdot{}\frac{\sqrt{x^2-1}}{2}\ln
  \frac{x+1}{x-1}\IEnn\\
  &&+\frac{i}{2}\sigma^{\mu\nu}\mathrm{P}_L
  \frac{p_{\nu}}{m_u}\frac{x-x_2}{x_1-x_2}
  \qty[\ln \frac{x+x_1}{x-x_2}-x_1\ln \frac{x+1}{x-1}]\IEnn\\
  &&+\frac{i}{2}\sigma^{\mu\nu}\mathrm{P}_R
  \frac{p_{\nu}}{m_b}\frac{x+x_1}{x_1-x_2}
  \qty[x_2\ln \frac{x+1}{x-1}-\ln \frac{x+x_1}{x-x_2}]\IEnn\\
  &&+\mathrm{P}_L \frac{p^{\mu}}{m_u}\frac{x-x_2}{x_1-x_2}
  \biggl[-2 +\qty(\frac{3}{2}+\frac{x_1+x_2}{x_1-x_2})\ln
  \frac{x+x_1}{x-x_2}\IEnn\\
  &&\hspace{8.7em}+\qty(\frac{2}{x_1-x_2}+x+\frac{3}{2}x_1)\ln
  \frac{x+1}{x-1}\biggr]\IEnn\\
  &&+\mathrm{P}_R \frac{p^{\mu}}{m_b}\frac{x+x_1}{x_1-x_2}
  \biggl[ 2+\qty(\frac{3}{2}-\frac{x_1+x_2}{x_1-x_2})\ln
  \frac{x+x_1}{x-x_2}\IEnn\\
  &&\hspace{8.7em}+\qty(\frac{2}{x_1-x_2}+x-\frac{3}{2}x_2)\ln
  \frac{x+1}{x-1}\biggr]
  \label{eq:C.2}
\end{IEEEeqnarray}
and the finite part $F(x,x_1,x_2)$ is
\begin{IEEEeqnarray}{rCl}
  F(x,x_1,x_2)&=&\frac{x}{2}\Biggl[\qty(3-2\ln
  \frac{\sqrt{\qty(x_1^2-1)\qty(x_2^2-1)}}{x^2-1})\ln
  \frac{x+1}{x-1}-\qty(\ln \frac{x_1+1}{x+x_1})^2\IEnn\\
  &&\hspace{2em}-\qty(\ln \frac{x_2+1}{x_1-x_2})^2+\qty(\ln
  \frac{x_1+1}{x_1-x_2})^2+\qty(\ln \frac{x_2+1}{x_1-x_2})^2\IEnn\\
  &&\hspace{2em}-2\ln \frac{x_1+1}{x+x_1}\ln
  \frac{x_1+1}{x_1-x_2}+2\ln \frac{x_2+1}{x_1-x_2}\ln
  \frac{x_2+1}{x-x_2}\IEnn\\
  &&\hspace{2em}+4\ln \frac{x_1-1}{x_2-1}\ln
  \frac{2}{x_1-x_2}+4\qty(\mathrm{Li}_2 \frac{1-x_1}{2}-\mathrm{Li}_2 \frac{1-x_2}{2})\Biggr]
 \label{eq:C.3}
\end{IEEEeqnarray}
In Eqs.(\ref{eq:C.2}) and (\ref{eq:C.3}), $\gma_{\mathrm{E}}$ is the
Euler-Mascheroni constant, $\mathrm{Li}_2(z)$ is the polylogarithm
function of order 2, and the definitions of $x,x_1,x_2$ are
\begin{IEEEeqnarray*}{rCl}
  x&=&\frac{m_b^2+m_u^2-p^2}{\sqrt{\kappa{}(m_b^2,m_u^2,p^2)}}\\
  x_1&=&\frac{m_b^2-m_u^2+p^2}{\sqrt{\kappa{}(m_b^2,m_u^2,p^2)}}\\
  x_2&=&\frac{m_b^2-m_u^2-p^2}{\sqrt{\kappa{}(m_b^2,m_u^2,p^2)}}
\end{IEEEeqnarray*}
where $\kappa{}(m_b^2,m_u^2,p^2)$ is the K{\"a}ll{\'e}n function or triangle function
\begin{equation*}
  \kappa{}(m_b^2,m_u^2,p^2)=\qty(m_b^2)^2+\qty(m_u^2)^2+\qty(p^2)^2
  -2m_b^2p^2-2m_u^2p^2-2m_b^2m_u^2.
\end{equation*}

The renormalization constant $Z_2^{b,\bar{u}}$ in Eq.\eqref{eq:41} has been computed in the virtial gluon-mass
regularization scheme and with the on-shell renormalization condition. The result is
\begin{equation}
  Z_2^{b,\bar{u}}=1+\frac{\alpha_s C_F}{4\pi}\qty(-\frao{\varepsilon}
  +\gma_{\mathrm{E}}-\ln \frac{4\pi\mu^2}{m_{b,\bar{u}}^2}-4
  -2\ln \frac{\lambda^2}{m_{b,\bar{u}}^2})
\end{equation}

\acknowledgments{}

This work is supported in part by the National Natural Science
Foundation of China under Contract No.11375088.

\bibliographystyle{apsrev4-1}

\end{document}